# Controlling magnetisation's reversal mechanism and hyperthermia efficiency in core/shell magnetic nanoparticles by tuning the interphase coupling

K. Simeonidis[1,2], C. Martinez-Boubeta[3,*], D. Serantes[4,5,*], S. Ruta[4], O. Chubykalo-Fesenko[6], R. Chantrell[4], J. Oró-Solé[7], Ll. Balcells[7], A.S. Kamzin[8], R.A. Nazipov[9], A. Makridis[1], M. Angelakeris[1,*]

[1]*Department of Physics, Aristotle University of Thessaloniki 54124, Greece*
[2]*Ecorecources P.C., Giannitson-Santaroza Str. 15-17, 54627 Thessaloniki, Greece*
[3]*Freelancer in Bilbao 48007, Spain*
[4]*Department of Physics, University of York, Heslington, York YO10 5DD, United Kingdom*
[5]*Applied Physics Department and IIT, Universidade de Santiago de Compostela 15782, Spain*
[6]*Instituto de Ciencia de Materiales de Madrid, CSIC, Cantoblanco 28049, Spain*
[7]*Institut de Ciència de Materials de Barcelona, CSIC, Bellaterra 08193, Spain*
[8]*Ioffe Physical-Technical Institute, Russian Academy of Sciences, St. Petersburg, Russia*
[9]*Kazan National Research Technological University, Kazan, Russia*

*Corresponding author: cboubeta@gmail.com, david.serantes@usc.es, agelaker@auth.gr

**Abstract**

Magnetic particle hyperthermia, in which colloidal nanostructures are exposed to an alternating magnetic field, is a promising approach to cancer therapy. Unfortunately, the clinical efficacy of hyperthermia has not yet been optimized. Consequently, routes to improve magnetic particle hyperthermia such as designing hybrid structures comprised from different phase materials are actively pursued. Here we demonstrate enhanced hyperthermia efficiency in relative large spherical Fe/Fe-oxide core/shell nanoparticles through the manipulation of interactions between the core and shell phases. Experimental results on exemplary samples with diameters in the range 30-80 nm indicated a direct correlation of hysteresis losses to the observed temperature elevation rate with a maximum efficiency of around 0.9 kW/g. The absolute particle size, the core/shell ratio, and the interposition of a thin wüstite interlayer, are shown to have powerful effects on the specific absorption rate. By comparing our measurements to micromagnetic calculations we have unveiled topologically non-trivial magnetisation reversal modes under which interparticle interactions become negligible, aggregates formation is minimized, and the energy that is converted into heat is increased. This information has been overlooked till date and is in stark contrast to the existing knowledge on homogeneous particles.

**INTRODUCTION**

Synchronization in natural processes, usually emerges spontaneously as long as there is some form of coupling between units in an oscillatory network. In magnetic nanoparticles, performance is defined by the non-monotonic behaviour of the collectivity rather than the summed individual contributions of each magnetic unit. One paradigm is the capability of magnetic nanoparticles to convert radiofrequency (RF) electromagnetic energy into heat. This has drawn a strong and growing research interest during the past decade, with examples in the field of catalysis,[1] lightweight thermoplastic composites for aeronautical and automotive engineering,[2] and numerous biomedical applications as well.[3,4] In particular, biocompatible magnetic nanoparticles have emerged as promising agents for use in drug delivery[5] and selective destruction of tumours by hyperthermia.[6] Numerous clinical and basic studies have shown that temperature stresses can alter tumour endurance in a significant manner. The use of RF fields instead of conventional methods allows for minimally invasive and real-time control of the temperature, particularly in deep body regions, while preventing damage to healthy tissues. At present, many fundamental research studies have been carried out worldwide and some have entered into clinical trials, especially when magnetic hyperthermia is combined with more traditional therapeutic approaches such as the co-delivery of anticancer drugs or radiation therapy.[7]

Despite rapid progress, the precise way to increase the heating efficacy of magnetic nanoparticle-based therapeutics is unknown.[8] This has posed a challenge for theoretical modelling. In fact, most of the reported hyperthermia data is encompassed by a universal response.[9] On the one hand, manipulation of the field conditions is the primary source of efficiency control.[10,11] Importantly, recent numerical analyses carried out by using 3D realistic models of the human body indicated that acceptable values for the magnetic field amplitude/frequency product (H·f) may increase up to four times greater than the usually considered safety threshold of ~5 ×10$^8$ A/ms.[12] On the other hand, factors such as the nanoparticle features (composition, size, shape), their geometrical arrangement and volume concentration within the dielectric cellular matrix also play an important role.[13] And so, in practice, the heat transfer from magnetic nanoparticles subjected to an oscillating field is not only affected by their intrinsic properties but also by the surrounding environment. For instance, the thermal conductivity of nanofluids is largely dependent on whether the nanoparticles stay dispersed, form large aggregates, or assume a percolating linear configuration.[14]

Minimizing the dosage of nanoparticles required for an effective treatment is another task of high significance. In that respect, ferrite nanoparticles are undoubtedly the most convenient building blocks for magnetic hyperthermia due to their good biocompatibility and stability. But conventional single-phase iron oxide systems suffer from low heating efficiency. Any further improvement of their known performance would involve precise chemical engineering on the nanoscale and interface structure.[6,15] Here, we present results of an experimental study that directly maps the structural and magnetic properties of iron-based core-shell particles. In particular, it is shown that surrounding a ferromagnetic (FM) iron core with a ferromagnetic iron oxide shell can tune the specific absorption rate (SAR) to unprecedented efficiency. Additional modellisation demonstrates different reversal mechanisms depending on the core/shell ratio and absolute particle size. More crucial to the synchronized behaviour of this multiphase material, we find that the presence (or not) of a nonmagnetic interlayer, which may completely change the interphase coupling and reversal mechanism, is key to allowing for diminished importance of the detrimental inter-particle dipolar interactions. This constitutes a remarkably different approach in comparison with other works dealing with core/shell geometries in which the objective was essentially to tune the effective single-particle properties but still under the giant-spin (i.e. coherent rotation of the entire system) hypothesis.[16,17] This scenario opens the way to a superior control of the heating performance, particularly for *in vivo* applications where the particles gradually attenuate their records.

**RESULTS AND DISCUSSION**
**Structure, composition and magnetism**
Compositional tuning is a major strategy to tailor magnetic properties. To this end, we have fabricated a series of complex structures using solar vapour phase condensation by varying the Fe-to-Fe₃O₄ target ratio, the chamber pressure and the oxidizing conditions as defined by pumping gas. Briefly, the evaporation of Fe under inert conditions preserves the metallic characteristics in a core that is lately passivated with a thin oxide shell, while simultaneous Fe$_3$O$_4$ decomposition also produces a small metallic core but surrounded by a thicker oxide shell. Moreover, depending on the composition of the gas stream during synthesis (Ar and/or air), completely oxidized nanoparticles (γ-Fe$_2$O$_3$) may also be produced.[18] Table 1 shows a compilation of both the preparation conditions, structural and morphological results extracted from a compendium of characterisation techniques. Figure S1 (see Supporting Information) depicts SEM imaging of samples under study.

**Table 1.** Overview of preparation conditions, main characteristics and schematic illustrations of studied samples.

| Sample | Target (% wt.) Fe | Target (% wt.) Fe$_3$O$_4$ | Chamber's pressure (torr)/Gas | Composition (% wt.) XRD/Mossbauer analysis Fe | FeO | Fe$_3$O$_4$ | γ-Fe$_2$O$_3$ | Average size (nm) | Morphology* Core/Shell(s) (nm) | Scheme |
|---|---|---|---|---|---|---|---|---|---|---|
| F01 | 100 | - | 70/Ar | 79.4 <br> 77.7 | - | 20.6 <br> 22.3 | - | 50 | 44/3 | 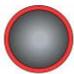 |
| F02 | 75 | 25 | 70/Ar | 58.6 <br> 69.8 | 10.1 <br> 5.2 | 31.3 <br> 25.0 | - | 52 | 42/2/3 | 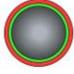 |
| F03 | 75 | 25 | 50/Ar | 53.6 <br> 67.5 | 11.8 <br> 13.1 | 34.6 <br> - | - <br> 19.3 | 33 | 25/1/3 | 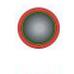 |
| F04 | 50 | 50 | 70/Ar | 12.9 <br> 11.1 | 14.7 <br> 8.4 | 72.4 <br> 80.6 | - | 48 | 24/4/8 | 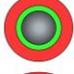 |
| F05 | - | 100 | 80/Ar | 9.1 <br> 5.9 | - | 90.9 <br> 94.1 | - | 78 | 30/24 | 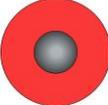 |
| F06 | - | 100 | 80/Ar-O$_2$ | - | - <br> 4.6 | - <br> 58.0 | 100 <br> 37.4 | 43 | 43/- | 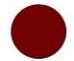 |

*estimated by geometrical features, compositional analysis and magnetic properties

In general, our experimental system is a nanocrystalline magnetic sphere composed of several layers (up to three) of Fe and its oxides. The initial structural analysis of obtained nanopowders was performed by XRD diagrams (see Figure S2 and S3 in Supporting Information). A first set of samples (F01-02) includes a Fe grain size about 40 nm surrounded by variable oxides ~ 5 nm thickness. Results suggest that, when a metallic target is evaporated under inert gas flow (sample F01), a large percentage of Fe (up to 80 %wt.) is preserved in the prepared nanoparticles while the surrounding outer shell is probably related with the native oxidation. Another series (F03-05) comprises a constant Fe core (25 nm in diameter) coated by oxide shells with increasing thicknesses from 4 up to 24 nm. By introducing Fe$_3$O$_4$ in the evaporating pellet material, the oxide with respect to Fe content gradually increases and becomes the dominant phase when the Fe/Fe$_3$O$_4$ ratio in the target reaches 1:1 (sample F04). However, even in the evaporation of pure Fe$_3$O$_4$ a small percentage of Fe is still observed as a result of Fe$_3$O$_4$ decomposition during the evaporation procedure (sample F05). Finally, a pure oxide sample (F06) stands for comparison to the 40 nm metallic nuclei.

Further insight into the microstructure by using TEM and elemental mapping provides evidence about the core/shell distribution of Fe and its oxides, depicted in Figure 1 representatively, for the smallest overall diameter (sample F03). Low resolution images (Figure 1a) clearly show the morphology of the particles and point to the presence of a lower contrast material casing a darker centre, but the presence of the core/shell morphology is definitively proven by the high-resolution TEM image of the individual particles. The electron diffraction results (Figure 1b), in consistence to XRD, suggest the coexistence of pure metal and oxide. Figure 1c presents a high contrast difference of the core and shell regions where the phase is indexed as Fe$_3$O$_4$. Further, local electron energy loss spectra analysis across the whole particle (Figure 1d) confirms that element distribution corresponds to the iron core/oxide shell morphology.

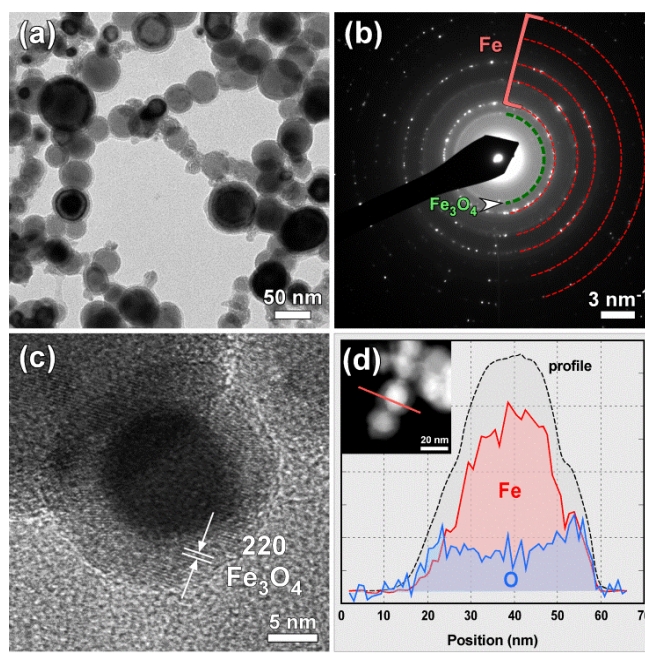

**Figure 1.** Representative TEM image of sample F03 (a), and corresponding selected area electron diffraction pattern (b) showing co-existence of metallic iron (red rings) and magnetite (green ring). Focusing on a small particle, HRTEM discloses a fluffy coating layer of magnetite (c). STEM mode imaging and composition profile derived from energy dispersive spectroscopy analysis demonstrating the iron-rich core (d).

We note that, independently of the median size, the relative distribution width is constant at about 20 %, a characteristic of our synthesis process.[18,19] Indeed, it is obvious that in real situations there are spreads in the volume of particles but also the system may contain a number of different phases, each with a specific saturation magnetic moment and anisotropy energy, and, therefore, several relaxation parameters. Mössbauer spectroscopy measures directly the contribution from magnetically distinct phases. At room temperature (see Table S1 and Figure S4 in the Supporting Information) the observed intensity ratio 3:2:1 between the sextet lines and the value of the hyperfine magnetic field (≈ 33 T) indicate that the core of most of the samples (except F06) is composed of bcc Fe, together with broad oxide components superimposed to the metallic iron. This picture is consistent also with the XRD and electron microscopy analyses.

Within the resolution of the structural analysis, it appears that the magnetic state of the nucleus remains unchanged, no matter the shell. In this regard, the iron oxide phases consist of two components ($Fe_3O_4$ and $Fe_2O_3$-like). We and others have demonstrated that any sample made of nanocrystalline iron-oxides is a mixture of several stoichiometries, whereas the small particles are maghemite rich.[20] It is well-established that the octahedral $Fe^{2+}$ ions located at the surface of magnetite particles are very sensitive to oxidation,[21] thus giving rise to this $Fe^{3+}$ rich layer, more visible in the smaller particles (sample F03) and those grown under the oxidizing atmosphere (sample F06). The corresponding oxide phase (mostly $Fe_3O_4$) in the case of pure iron targets (sample F01) is explained by the inevitable surface passivation after exposure to ambient conditions. Moreover, every time that the target was a mixture of Fe and $Fe_3O_4$ in variable proportions (samples F02, F03, F04), the Mössbauer spectrum shows also a broad paramagnetic central peak, which can be assigned to nonstoichiometric wüstite (FeO is an antiferromagnet with a Néel temperature around 200 K). It is worth mentioning that FeO is known to disproportionate into Fe and $Fe_3O_4$.[22] Giving a model to explain the oxidation process of iron, it seems reasonable that this ferrous FeO phase would be placed at the core/shell interface, between the pure metallic Fe and the oxidized outer region.[23,24] Obviously, epitaxial strains arise due to the accommodation of the crystalline lattices of the oxide and the metal. And probably, this strain favours the nucleation of the otherwise metastable FeO-like phase. For example, in thin films, the misfit between the iron atoms in the [-110] direction of FeO and the [001] direction of iron is approximately 6%, whereas between FeO and $Fe_3O_4$ it is only 3%. Regarding the oxidation of iron nanoparticles, there is still little information on the evolution of growth stresses in these multiphase regimes. Naturally, a tensile stress is expected in the $Fe_3O_4$ shell opposing the compressive strain on the core,[25] and this stress increases, at the very initial phase, with the thickness of the oxide layer.[26] In our case, stress occurrence as witnessed by XRD main peak shift attributed to reduction in average lattice parameter is reported in the inset of Figure S2 and Figure S3 within the Supporting Information.

Magnetization measurements shown in Figure 2 demonstrate different reversal mechanisms depending on the core/shell ratio and absolute particle size. On the one hand, the saturation magnetisation increases with the percentage of iron, and reaches 200 $Am^2/kg$ for F01, which is close to the pure bulk value. But, as shown in Figure 3, this trend is not reflected in monotonic behaviour of the Mr/Ms ratio and coercivity as Fe content decreases and oxide percentage is added gradually to the shell. The coercivity first increases, reaches a maximum at a Fe-to-oxide ratio of 2 and then decreases. We highlight that for iron, maximum coercivities due to the magneto-crystalline anisotropy alone are about 40 mT,[27] though strain anisotropies may force it up to 60 mT, which is the case reported here. Complete oxidation of nanoparticles (sample F06) results again in weaker magnetic features.

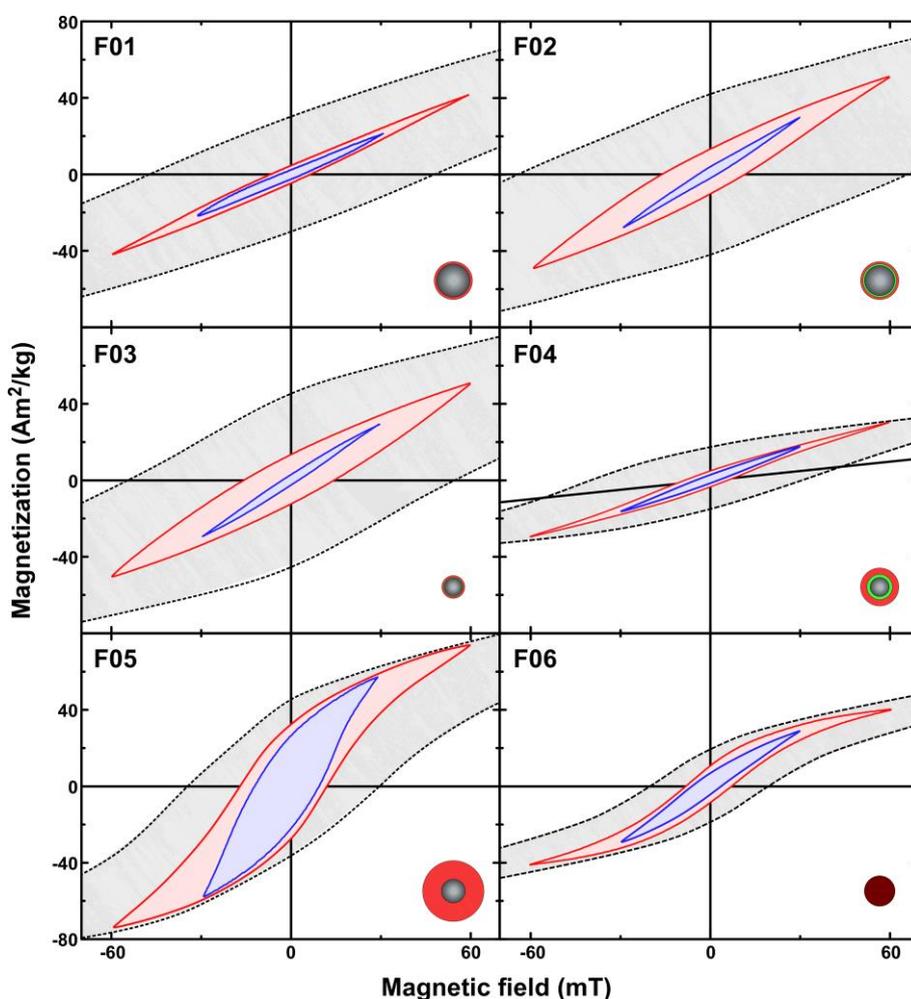

**Figure 2.** Minor hysteresis loops obtained by cycling between lower peak-field values of 30 (in blue) and 60 (red) mT for samples under study with decreasing Fe content (F01-F06). The grey-shaded region accounts for the corresponding full cycles at 1 T, well above saturation field.

Incidentally, room-temperature magnetic properties of core/shell iron/iron oxide nanocrystals with core sizes ranging from ~ 5 to 20 nm, and constant thickness of the oxide shell about 2 nm, were previously investigated by Kaur *et al*.[28] They concluded that dipolar interparticle interactions become

-5-

stronger with the growth of size. There, the decrease in $H_c$ from 50 mT to about 5 mT with decreasing particle size was due to thermal effects. This is not the case here, and we surmise that not only interparticle but also intraparticle interactions between the iron core (ferromagnetic) and oxide shells should be considered.

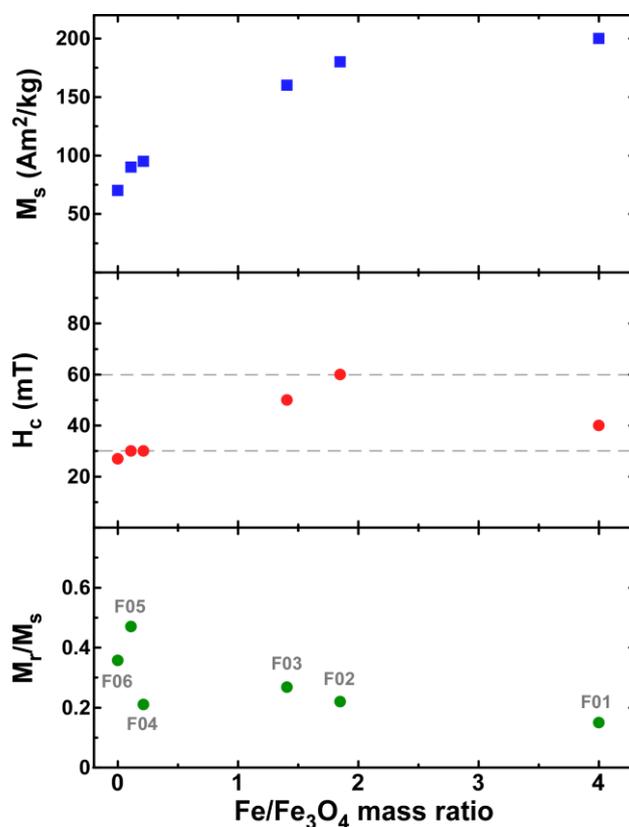

**Figure 3.** Room-temperature dependence of saturation magnetisation, coercive field and remanence ratio on the $Fe/Fe_3O_4$ mass ratio of core/shell magnetic nanoparticles. Horizontal lines in the $H_c$ plot account for maximum fields in minor cycles (30 and 60 mT) and hyperthermia characterization.

So far we have determined the quasi-static magnetic response in a core/shell iron/iron oxide system. A combined study of microscopy, XRD and Mössbauer reveals the existence of several oxides in the shell region, with a major contribution from $FeO$ and $Fe_3O_4$. Apart from the compositions, hysteresis loops demonstrate that optimizing material parameters deals with core/shell size proportion. Thus, it is anticipated that particular ratios of soft to hard phases would lead to enhanced dynamic response.

**Magnetic heating**

The efficiency of magnetoheating particles is primarily governed by three interrelated parameters: the magnetic anisotropy energy (which depends on several features such as the size of the particles), the experimental field conditions, and the dosage (see Supporting Figures S6 and S7). Figure 4 indicates that since the amount of heat generated is given by the product of the frequency and the area of the hysteresis loop, SAR is gradually enhanced upon increasing the frequency. Concomitantly, according to the hysteresis losses (see Figure 2), oxide-rich nanoparticles are by far more efficient at lower AC fields (see data at 30 mT) while Fe-rich samples require higher fields to unfold their full response, as also anticipated from the $H_c$ and $M_r$ dependencies shown in Figure 3. On the other hand, the complete oxidation results to similar response than that for typical iron oxides reported in literature.[29] Thus, it seems that tuning of the two phases is in favour of both magnetic hardening together with enhanced heating efficiency provided the proper hyperthermia conditions are applied (AC frequency and field amplitude). Indeed, we acknowledge that similar (bi)magnetic iron/iron oxide core/shell nanoparticles (though smaller < 15 nm in size) have been already explored for *in vivo* hyperthermia.[30] Interestingly, in our case, sample F05 with oxide shell thickness ($t_S$ = 24 nm) and Fe core's radius ($r_C$ = 15 nm), shows a remarkable heating efficiency which approaches 0.9 kW/g. On this, one should consider the coincidence that the term $(t_S+r_C)/t_S$ is found practically equal to the term $t_S/r_C$ and both very close to the value of φ = 1.618... which is known as the geometric golden ratio and thus suggesting that nanoscale effects may be also governed by the conception of symmetry and perfection which is met in many instances of nature, science and human art.

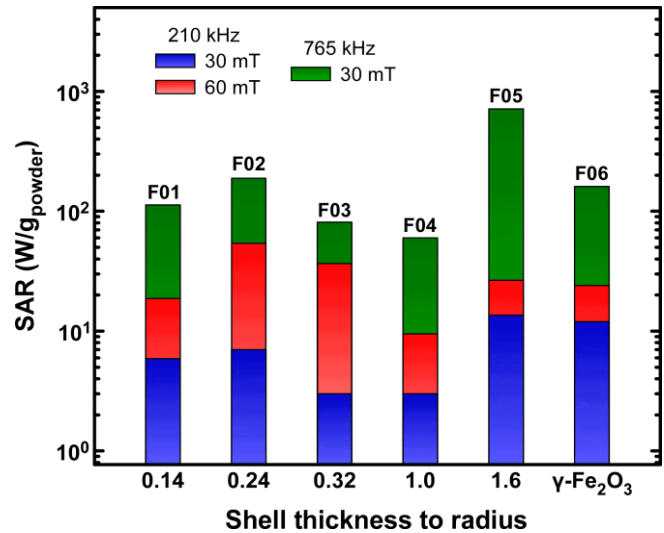

**Figure 4.** Hyperthermia efficiency as a function of fixed frequency (210 kHz) but different field amplitude, and the same field amplitude (30 mT) but different AC field frequency. Particles concentration is 4 mg/mL. Sample F06 would correspond to infinite ratio.

**Magnetic modelling**

The simplest and most generally employed model of magnetic hyperthermia is the linear response theory of Rosensweig.[31] For a given particle size this gives a scaling of the heating with $fH^2$. The measurements given in Figure 4 clearly do not conform to this scaling. To explain the behaviour of such complex systems one must use a model allowing the possibility of non-uniform reversal with the capability to introduce the detailed structure of the core-shell combination. In an effort to infer which are the main core/shell features determining the magnetic response (e.g. aspect ratio, total size, etc.) and to correlate those with the heat release, a theoretical approach based on a micromagnetic technique using the OOMMF software package[32] was carried out. An illustrative example of the micromagnetic simulations is shown in Figure 5. The simulated structure has a Fe core of radius $r_C$, a magnetite shell of thickness $t_S$, and a FeO interlayer of width $t_{sp}$ treated as paramagnetic. For the sake of simplicity we have focused on one field amplitude and frequency case with $H_{max}$ = 30 mT and f = 765 kHz.

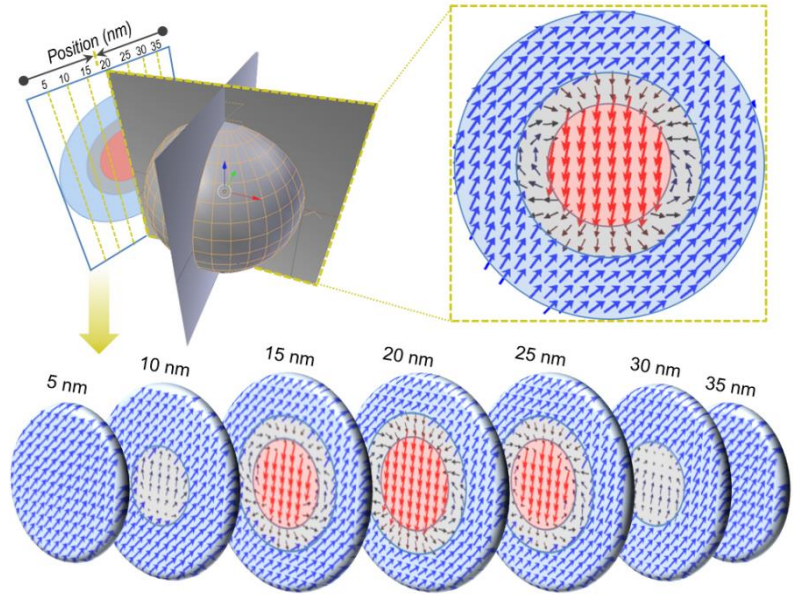

**Figure 5**. Micromagnetic magnetisation configuration at remanence of an example core/shell structure ($D_{tot}$ = 40 nm; $r_C$ = 16 nm, $t_{sp}$ = 4 nm; $t_S$ = 8 nm), in this case showing AFM-like coupling between core and shell. The snapshots are taken at regular cross-sections of the particle, as illustrated by the sketch shown up, left. For clarity purposes the arrows representing the magnetisation are magnified (each arrow corresponds to 8 cubic subunits of 1-nm side each).

Based on the experimental details reported in Table 1, some representative examples of computed hysteresis loops are summarized in Figure 6, illustrating the different roles of the particle size $D_{tot}$ (panel A), the core/shell $t_S/r_C$ ratio (B); thickness of the paramagnetic interlayer spacer (C); and average size of shell crystallites (D), on the magnetisation response. We have focused on one field-amplitude/frequency case, $H_{max}$ = 30 mT and f = 765 kHz, which based on the large sizes of the particles considered we can reasonably assume corresponds to FM-like behaviour. It is important to recall these are just examples corresponding to some specific particle characteristics; exhaustive analysis is reported in the Supporting Information under the ***S.3 Core/shell parameter dependences*** section. Overall, the results suggest several key roles of the four parameters outlined above, namely:

- A. Total particle size: for small values of $D_{tot}$ the particle behaves as a large magnetic *supermoment* with coherent magnetisation behaviour, while increasing $D_{tot}$ leads to a non-coherent magnetisation reversal and narrower hysteresis loops. Interestingly, the threshold between regimes – which corresponds to an enhanced loop area- occurs at the transition at which the exchange energy is on equal footing to the magnetostatic one.
- B. Core/shell ratio: for a given $D_{tot}$, magnetisation switching evolves from a non-coherent behaviour (of small hysteresis area values) to a fully coherent one (with associated large areas) with increasing the $t_S/r_C$ value. Interestingly, the transition occurs around $t_S/r_C$ = 1.
- C. Thickness of interlayer spacer: the presence of a PM-like interlayer allows antiferromagnetic (AFM) coupling between the core and shell, being the soft shell the dominating phase, i.e. forcing the core to switch and also polarising the FeO. In such cases the hysteresis area may be significantly enlarged.
- D. Polycrystalline shell may lead to a complete change in the shape of the loop, from smooth to very abrupt (squared) that would be highly desired for enhancing the heat output.[33] Worth to note, in this case both Fe and $Fe_3O_4$ phases are FM-like coupled and the magnetisation reversal is essentially coherent.

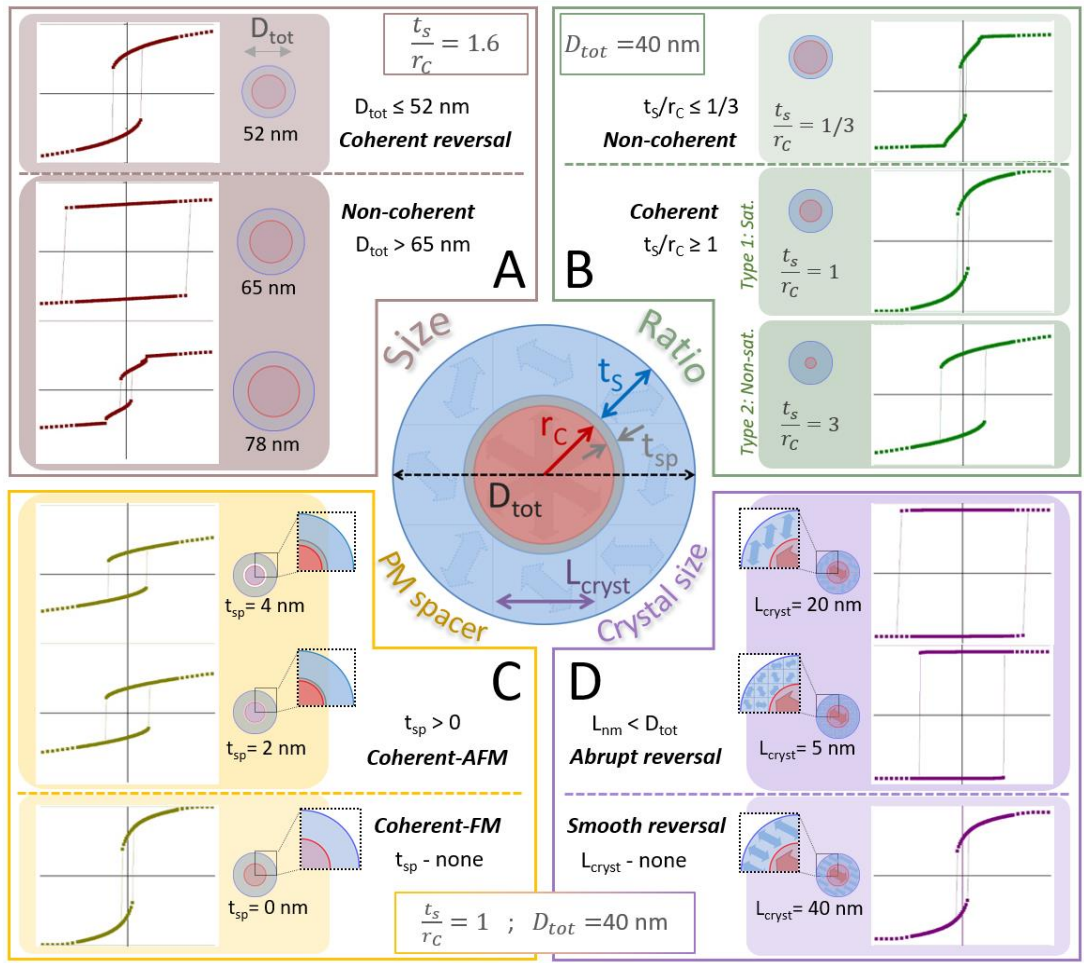

**Figure 6.** Case study of the hysteresis loops for an isolated particle depending on the total particle size (A); core/shell ratio (B); thickness $t_{sp}$ of the paramagnetic interlayer spacer (C); and average size $L_{cryst}$ of crystallites composing the shell (D). The central sketch depicts the associated core/shell structure.

In addition, another important characteristic of the presence of a polycrystalline shell is that it significantly influences the dependence of the magnetic response on the direction of field application (see the discussion under the Supporting Information section ***S.4 Angular-dependent properties and polycrystalline shell***). This is an important aspect regarding the generality of the results, since the reported trends correspond, unless otherwise stated, to a particular configuration of applied magnetic field (along the (1 0 0) direction, parallel to an easy axis of the Fe core). The comparison between experimentally measured and theoretically predicted heating performance is shown in Figure 7. Very good agreement between experiment and theory is obtained,[34] except in the case of sample F03. Additional modelling (see the Supporting Information section ***S.5 Simulation of the experimental samples***) demonstrates that there is a crucial difference between sample F03 and rest: sample F03 is the only one not completely reversing the magnetisation, i.e. undergoing minor loop conditions – may this feature be enough to justify the divergence?

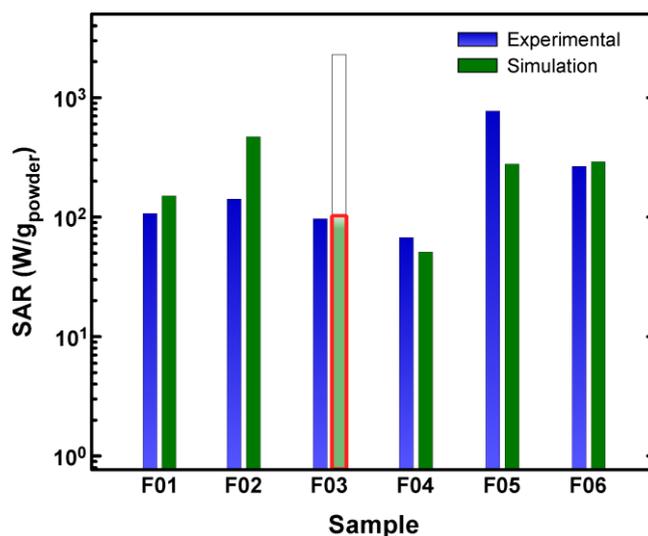

**Figure 7.** Comparison between experimentally measured SAR values (blue bars) and the theoretical predictions (green bars) obtained based on the model reported in Figure 6, for non-interacting conditions and field characteristics of f = 765 kHz and $H_{max}$= 30 mT. The SAR values are estimated from the area of the loops as SAR=f*Area. Note that the computed heating efficiency for sample F03 describes the observational studies only if particle-particle dipolar coupling is included (marked red column), further explained in the text.

The implications of such minor-loop conditions regarding heating performance can be extremely important (minor loops correspond to much smaller heating power than the maximum achievable one; not to mention other effects such as e.g. higher dispersion in local heating performance[35]). However, the fact is that experimentally the heat dissipated by this sample is not negligible, but comparable to those of samples F01 and F02 - and even higher than that of F04. Clearly, there is significant heat dissipation from sample F03 that is not predicted by the theoretical model. Even if a random distribution is computed, the predicted heat release becomes significantly higher than the experimentally measured value, as illustrated by the green shaded tall column of sample F03 within Figure 7 (see Supporting Figure S26 and discussion therein).

To investigate this complex scenario, we have performed various modifications to the model, including the addition of perpendicular surface anisotropy in the outer layer of the shell and/or at the interface, dynamical effects; etc. The results, summarized in the Supporting information section ***S.6 Trial model modifications***, indicate no improvement in the overall experiment-theory comparison but, on the contrary, increased divergence in some other samples. So we are left with the conclusion that *the presumed hypotheses stating that interparticle interactions plays a secondary role does not hold for sample F03*, and that the dipole-dipole coupling changes the field parameters necessary to achieve the maximum SAR.[36] Our assumption is that non-reversing particles in the minor condition create a static field on the reversing particles. In order to test this, we next simulated the heating performance of pairs of potentially interacting particles from sample F03 at several separations and arrangements (see Figure S30 within the Supporting information). The results indicate a decrease of the heating performance with decreasing interparticle distances, and only if we allow for some form of coupling to emerge at less than 1.5 times the particles' diameter can we meaningfully explain the measured SAR values.

The reason why this effect of particle-particle interactions on SAR is only observed in sample F03 stems from the fact that this is the only sample exhibiting both requirements of coherent reversal and minor loop behaviour. As a complementary source of information, Figure 8A compares the magnetization at remanence and Figure 8B compares the corresponding stray fields of each particle. In particular, F03 clearly stands out from the other samples, having a strong sensitivity to magnetostatic interactions.

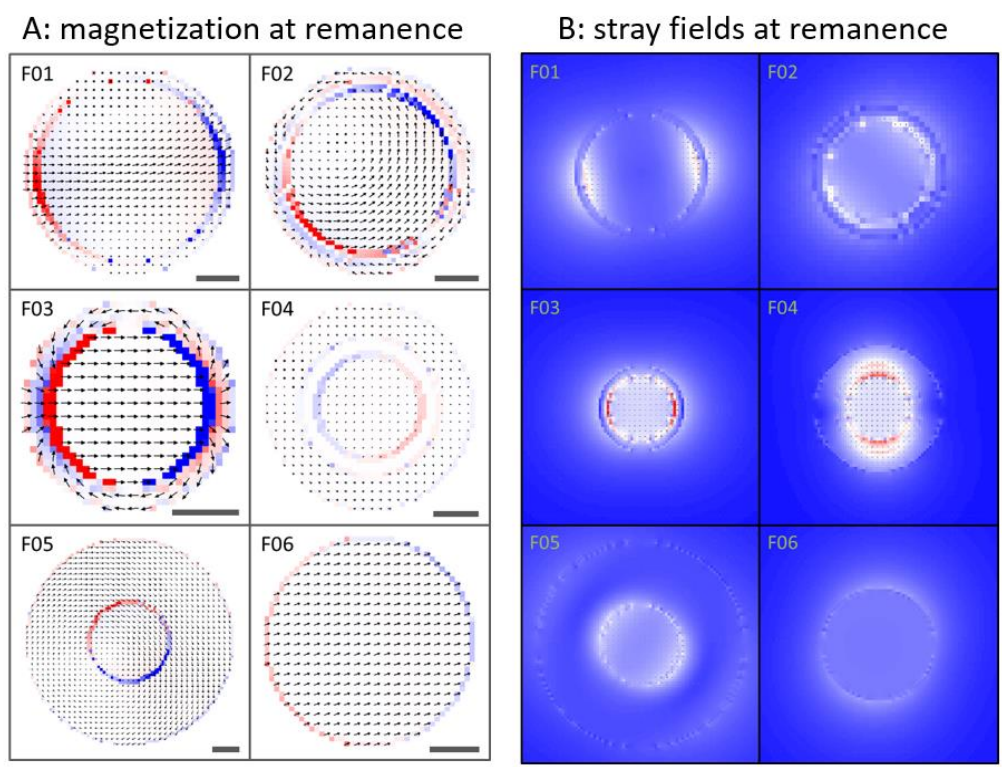

**Figure 8**. (A) Magnetisation configurations at remanence for all the samples discussed in Table 1, where the magnetisation (arrows; each one corresponds to the average over 8 unit cells) is superimposed with the magnetisation divergence (colours), which stands for the coherence and magnitude of the magnetic moment of the particle. The bars correspond to a 10 nm scale, and serve to illustrate the difference in sizes between the particles. (B) Demagnetizing fields at remanence. The arrows (each one corresponds to the average over 8 unit cells) show the field direction and the colours the magnitude of the demagnetizing field (stray field outside of the particles) from blue (zero field) to red (maximum value of 1168 kA/m). Note that in this case all particles are in the same scale bar; the common case between panels A and B is sample F05, the biggest.

Curved topology in magnetism attracts intensive research given its potential applications to information technology. Such non-uniform spin configurations have been also exploited by other researchers in the effort to fight cancer. For instance, the magnetic vortex in planar thin-film microdiscs creates an oscillation when a low frequency (tens of Hz) AC magnetic field is applied.[37] This movement transmits a mechanical force to the surrounding tissue that may be enough to compromise integrity of the cellular membrane. Additionally, coupling the vortex eigenmode to MHz frequencies may provide a new and efficient means of energy absorption by, and emission from, magnetic nanoparticles.[38] However, from the standpoint of micro-magnetism, our study differs completely from those not only because they were based on soft-magnetic Permalloy, while our particles comprise a core and shell iron derivatives, but also the frequency range considered, as well. Instead, our case resembles the one reported by Liu *et al.*[39] in iron oxide nanorings, in which magnetisation is circumferential to the ring without stray fields. Minimization of stray fields is also crucial to reduce dipole–dipole interactions and agglomeration of magnetic particles in solution, supporting the argument that the oxide shell promotes decoupling of the iron nuclei. And so, this work adds to the growing body of knowledge showing the complex issue of physical interpretation of interaction effects in magnetic hyperthermia. Finally, it is worth to related our work with the question posed by Z. Nemati et al.[16], *are core/shell magnetic nanoparticles promising for hyperthermia*? Our answer is not only a clear yes but, furthermore, we emphasize the rich possibilities open by tuning the core/shell coupling of those structures for tailor made specific response applications.

**CONCLUSIONS**

In summary, distribution of iron and its oxides in well-defined core/shell nanoparticles is proven a crucial parameter when it comes to optimal magnetic properties in hyperthermia. The curvature of the underlying surface in core/shell nanoparticles leads to a coupling between the localized out-of-surface component with its delocalized in-surface structure. This stems from the fact that the geometry of a spherical shell prohibits an existence of spatially homogeneous magnetisation, and when made of a magnetically soft ferro(i)magnet, for each thickness there exists a critical radius such that an onion-like state develops into a whirligig state. In our case, samples with non-coherent reversal seem more adequate in generating large SAR values because interparticle interactions are negligible in comparison with the inner magnetisation processes, thus, diminishing the possibility of aggregated formation. Combined experimental and theoretical analysis also revealed a novel magnetization process wherein the iron-oxides effectively reduce the coercivity of the ferromagnetic cores by leading the magnetization process at small magnetic fields, thus overcoming the main drawbacks of stand-alone homogeneous nanoparticles for magnetoheating. In particular, observations of transverse (or vortex) domains stress the difference between core/shell magnetic constructions and homogeneous particles.

**EXPERIMENTAL SECTION**

**Particles' preparation.** A very simple, fast, green and cost-effective way by using solar vapour phase condensation. This technique allows the preparation of large quantities of nanoparticle dry powders presenting a narrow particle size distribution without purification steps.[40] Families of particles were prepared using almost all of the 3d metals, and many others, but here we will restrain the discussion to iron, which is considered to be an essential nutrient for cells that are dividing rapidly, such as in tumours, and also plays a vital role to perform various body functions. Particularly, evaporation of pure Fe powder pressurized in a pellet form, under Ar flow (70 torr) resulted in 50 nm nanoparticles. Under the same pressure, co-evaporation of Fe and $Fe_3O_4$ mixtures produces a particle diameter at the same level but the thickness of oxide on the surface is anticipated to increase as compared to the shell stabilized after natural oxidation. The decrease in particles diameter (down to about 35 nm) was achieved by using lower pressure (50 torr) whereas evaporation of a pure $Fe_3O_4$ target at 80 torr enables the production of larger nanoparticles (75 nm) with a very low zero-valent Fe content. The pumping of oxygen is sufficient to provide a completely oxidized dark-red product ($\gamma$-$Fe_2O_3$).

**Morphology and crystal structure.** Structural characterization was performed by X-ray diffraction (XRD) using a Rigaku Ultima+ powder diffractometer with Cu-K$_\alpha$ radiation. Transmission electron microscopy (TEM) images and selected area electron diffraction (SAED) patterns were obtained on a JEOL JEM-1210 operating at 120 kV. A number of samples were examined in more detail using High-resolution (HRTEM) and scanning transmission electron FEI Tecnai G2 F20 microscope (STEM) operating at 200 kV and equipped with EDAX element analysis system. Samples were prepared by dispersing the powders in ethanol. A small droplet of the suspension was placed on a holey carbon film supported on a copper grid. Scanning electron microscopy (SEM) micrographs of numerous samples were obtained in a Quanta 200 ESEM FEG FEI microscope.

**Mössbauer spectroscopy.** In order to give more accurate description of co-existing iron phases, the powders were analytically investigated by $^{57}$Fe Mössbauer spectrometry. The spectra were collected at room temperature in transmission geometry, using a conventional constant acceleration spectrometer operating in triangular wave mode and a $^{57}$Co source into rhodium matrix. The speed scale of the Doppler modulator and isomer shift calibration were performed using a 10 μm thick α-Fe foil. Mössbauer spectra were fitted using the Mosfit software, a least-square iteration program.[41] Hyperfine interaction parameters are denoted as follows: IS- for isomer shift (mm/s), QS- for quadrupolar shift (mm/s) and $H_{eff}$ for magnetic hyperfine field (T). Estimated errors ($\pm$ 5%) originate from the statistical inaccuracy given by the fitting program. Results concerning composition were compared and validated to the quantified data from the corresponding XRD diagrams using the Rietveld methodology.

**Magnetism.** Magnetic features were evaluated by room temperature vibrating sample magnetometer (VSM – 1.2H/CF/HT Oxford Instruments) in a magnetic field range $\pm$ 1 T. Additional minor hysteresis loops at specific magnetic fields were also recorded after demagnetising the samples, in order to compare to theoretically estimates of hysteresis losses per cycle.

**Calorimetric measurements.** The heating efficiency of nanoparticles dispersions in distilled water was evaluated in two different devices using a higher frequency (765 kHz) AC magnetic field peak of 30 mT and a lower frequency (210 kHz) setup capable of operating in 30 to 60 mT field amplitude range. Temperature was monitored by using a GaAs-based fiber optic probe immersed in a test tube containing 1mL of dispersion. The quantifiable index of heating efficiency, the specific loss power, was derived from the slope of the temperature versus time curve in non-adiabatic conditions with detailed modelling of the heat exchange with the surrounding environment.[18,42,43]

**Computational details.** Micromagnetic simulations in the macrospin approximation were carried out using the 3D version of the OOMMF package.[32] In first approximation all particles were assumed spherical with a single-crystal core to avoid considerations regarding preferential orientation or the likely additional anisotropy contribution[44]. M(H) hysteresis loops were simulated for several core/shell structures in terms of the $Fe_3O_4$ shell thickness-to-Fe core radius, i.e. $t_S/r_C$ ratio. Sometimes a separating FeO interlayer of width $t_{sp}$ was included. In all cases we considered a 1 nm-side cubic cell discretization, i.e. below the smaller specific exchange length, $l_{exch1}$ = 3.7 nm for the iron core and $l_{exch1}$ = 8.4 nm for the $Fe_3O_4$ shell (see Supporting Information). Field was set collinear to the [100] easy axis of magnetisation of iron. No thermal excitations were considered (because of being computationally unfeasible), thus rendering the hyperthermia results applicable only to FM-like nanoparticles (reasonable assumption for the large sizes considered[45,46]). The common parameters used to characterise all core/shell particles are: shown in Table 2, with the exchange between core and shell assumed to be 0.175·10$^{-11}$ J/m (i.e. about 10% of the average of both phases). The easy axes of the shell are randomly oriented from crystallite to crystallite, as explained later in further detail. The separating spacer is treated as paramagnetic-like, with a constant $M_S$ value: much smaller than the other phases (i.e. easily polarised); and with neither exchange, nor anisotropy.

Table 2. Summary of the parameters used for the simulations.

| | $M_S$ (emu/cm³) | $K_C$ (erg/cm³) | $A_{exch}$ (J/m) | Axis1 | Axis2 |
|---|---|---|---|---|---|
| Fe core | 1711 | 4.6·10$^5$ | 2.5·10$^{-11}$ | (1,0,0) | (0,1,0) |
| $Fe_3O_4$ shell | 477 | -1.1·10$^5$ | 1.0·10$^{-11}$ | | |
| FeO spacer | 1 | 0 | 0 | | |

The simulations were performed assuming quasistatic conditions (reasonable for FM-like behaviour, as discussed elsewhere[9,35]), and the precision criteria for the energy minimization evolver was set to $\hat{m} \times \vec{H} \times \hat{m} = 0.01$ (i.e. the highest recommended by the code developers), where $\hat{m}$ is the magnetisation unit vector within each discretization cell and $\vec{H}$ is the external magnetic field, applied along the (1 0 0) direction. The discretization was always the same, fixed as 1nm-side cubic cell. To discard undesired artificial effects of the initial configuration choice on the magnetic behaviour, the system was in cases initialised as a vortex state; nevertheless, it is worth to note that starting at either the saturated or random configurations did not result in appreciable differences.

*Acknowledgements*. Financial support of the SFERA II project – Transnational Access activities (EU 7th Framework Programme Grant Agreement nº 312643) is acknowledged as well as the use of PROMES facilities and its researchers/technology experts. We acknowledge the Centro de Supercomputación de Galicia (CESGA) for the computational resources. This research was partially supported by the Consellería de Educación Program for Development of a Strategic Grouping in Materials (AeMAT) at the Universidade de Santiago de Compostela (Grant No. ED431E2018/08, Xunta de Galicia). D.S. also acknowledges Xunta de Galicia for financial support under the I2C Plan. K.S. acknowledges Stavros Niarchos Foundation for financial support. C.M.B. would like to thank his wife, Marian, for sustain during the many rough draft versions of this paper.

*Supporting Information Available:* xxxx This material is available free of charge via the Internet at http://pubs.acs.org.

**Graphical abstract**

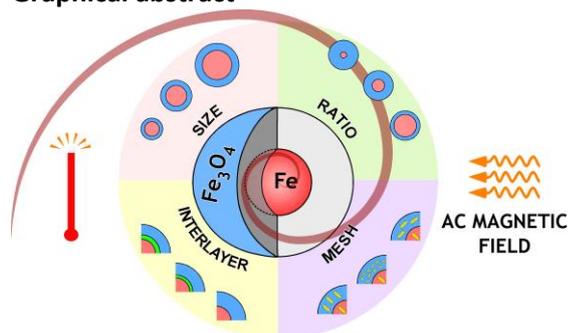

# Supporting Info

# Controlling the reversal mechanism and hyperthermia efficiency in core/shell magnetic nanoparticles by tuning the interphase coupling

K. Simeonidis, C. Martinez-Boubeta, D. Serantes, S. Ruta, O. Chubykalo-Fesenko, R. Chantrell, J. Oró-Solé, Ll. Balcells, A.S. Kamzin, R.A. Nazipov, A. Makridis, M. Angelakeris

We first present the morphology of the particles and probable crystal structure and, in a brief section, the main experimental observations on magnetic properties at room temperature and heating capabilities. We then use the simplest form of a finite difference simulation method to theoretically derive the main core/shell features determining the magnetic response (e.g. total size; phases proportion; etc.) and to correlate those with the heat release.

Figure S1 presents the SEM images of the prepared nanoparticles that were used to verify their spherical shape and estimate the mean average diameter. Particularly, evaporation of pure Fe powder pressurized in a pellet form, under Ar flow (70 torr) resulted in 50 nm nanoparticles (Sample F01: Figure S1). Under the same pressure, co-evaporation of Fe and $Fe_3O_4$ mixtures produces a particle diameter at the same level but the thickness of oxide on the surface is anticipated to increase as compared to the shell stabilized after natural oxidation (Samples F02-F03: Figures S1). The decrease in particles diameter (down to about 33 nm: Sample F03: Figure S1) was achieved by using lower pressure (50 torr) whereas evaporation of a pure $Fe_3O_4$ target at 80 torr enables the production of larger nanoparticles (78 nm: Sample F05: Figure S1) with a very low zero-valent Fe content. We recall the thermal conductivity of a gas is proportional to the mean free path and the gas density. Thus, in the vapor-condensation method, particle size generally increases with increasing the gas-pressure as a result of decreasing mean free path. Additionally, it was observed that the lighter the carrier gas (for instance, by partially replacing Ar with oxygen) the smaller the mean size of the particles (the case of F06 *vs*. F05: Figure S1).[1]

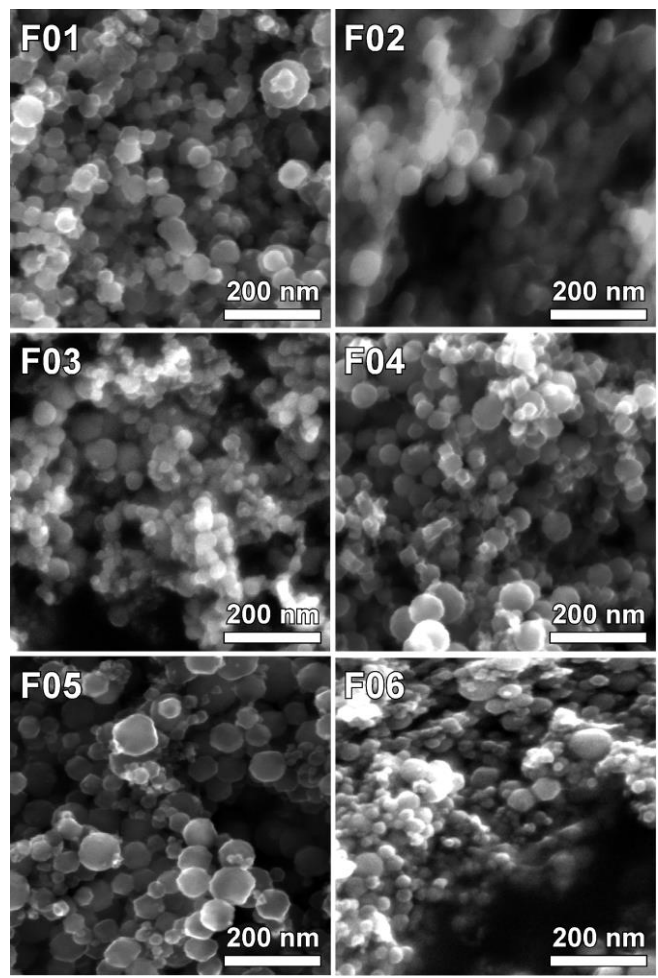

**Figure S1.** SEM images of representative samples from this study, in decreasing Fe content (F01-F06).

The experimental results of the morphological investigation and X-ray powder diffractometry suggest that crystals possess high quality. The patterns were fitted following the Rietveld methodology on the basis of bcc Fe, spinel oxides ($Fe_3O_4$, $\gamma$-$Fe_2O_3$) and FeO in order to get an indicative quantification of the observed structures.

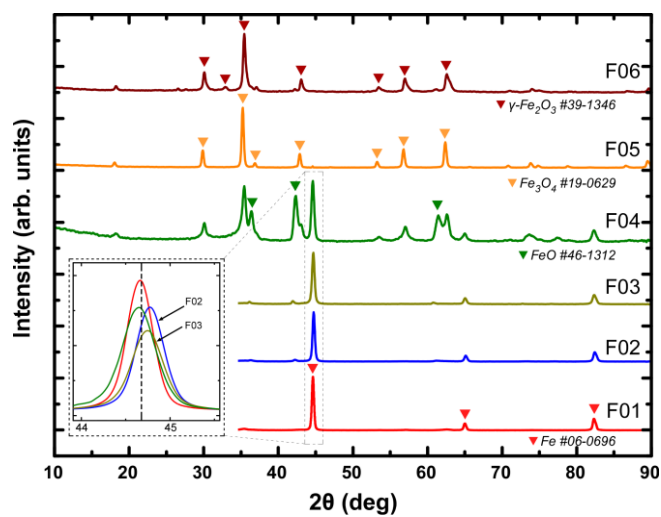

**Figure S2.** Comparative XRD patterns for samples under study. Inset presents a magnification of (110) diffraction peak suggesting the presence of epitaxial strain in the Fe core. We assume that a shift to higher angle in reciprocal space indicates compression, i.e. reduction in average lattice parameter.

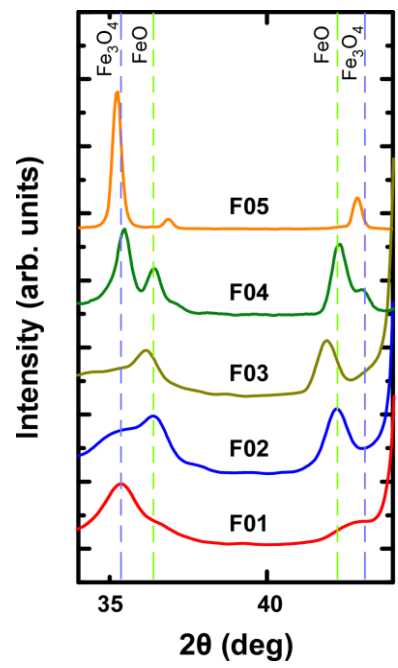

**Figure S3.** Magnification of XRD diagrams in the range of (311), (400) diffraction peaks for $Fe_3O_4$ and (111), (200) diffraction peaks for FeO.

Room temperature Mössbauer spectroscopy was used to differentiate the contribution from magnetically distinct phases. Table S1 summarizes the parameters estimated from the curves fitting. Relaxation models provided the best numerical fit,[2] with the very broad spectra lines as a consequence of distributions of the hyperfine parameters due to surface and interface disorder.[3]

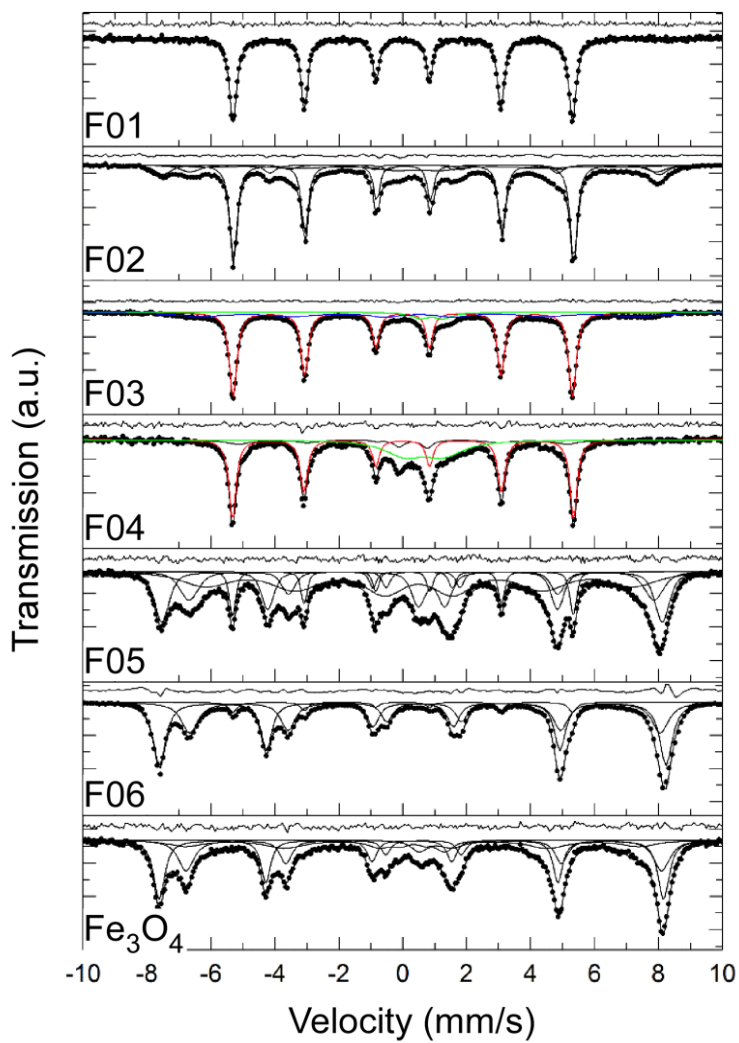

**Figure S4.** Mössbauer spectra of samples with varying Fe content: F01-F06 from top to bottom. The spectrum of the parental magnetite powder is used as a reference sample.

**Table S1.** Parameters obtained from experimental Mössbauer spectra. A and B account for sextets of iron cations in tetrahedral and octahedral oxygen coordination, respectively.

| Sample | Phase | Content (%) | IS(mm/s) | QS(mm/s) | $H_{eff}$(T) |
|---|---|---|---|---|---|
| **F01** | Fe | 77.7 | 0.03 | 0.00 | 33.0 |
| | $Fe_3O_4$ | 12.3 | A 0.31 | -0.08 | 48.3 |
| | | | B 0.67 | -0.05 | 45.3 |
| | RMS oxide ($Fe_3O_4$) | 6.1 | 0.47 | -0.04 | 40.9 |
| | | 3.9 | 0.47 | 1.37 | 0.0 |
| **F02** | Fe | 69.8 | 0.00 | 0.00 | 33.0 |
| | FeO | 5.2 | 0.92 | 0.74 | 0.0 |
| | RMS oxide($Fe_3O_4$) | 25.0 | 0.51 | 0.00 | 42.4 |
| **F03** | Fe | 50.8 | 0.00 | 0.00 | 33.2 |
| | Fe, nano | 16.7 | 0.01 | 0.02 | 32.0 |
| | FeO | 13.1 | 0.67 | 1.23 | 0.0 |
| | RMS oxide (Fe(III)) | 19.3 | 0.32 | 0.88 | 0.0 |
| **F04** | Fe | 11.1 | 0.00 | 0.00 | 33.1 |
| | $Fe_3O_4$ | 40.7 | A 0.30 | -0.02 | 48.6 |
| | | | B 0.62 | -0.05 | 45.1 |
| | FeO | 8.4 | 0.91 | 0.82 | 0.0 |
| | RMS oxide($Fe_3O_4$) | 39.9 | 0.51 | 0.02 | 41.5 |
| **F05** | Fe | 5.9 | 0.00 | 0.00 | 33.0 |
| | $Fe_3O_4$ | 94.1 | A 0.32 | -0.01 | 49.2 |
| | | | B 0.69 | 0.02 | 45.7 |
| **F06** | $Fe_3O_4$ | 58.0 | A 0.28 | -0.02 | 49.0 |
| | | | B 0.66 | 0.01 | 46.2 |
| | FeO | 4.6 | 0.92 | 0.81 | 0.0 |
| | RMS oxide(Fe(III)) | 34.3 | 0.36 | -0.12 | 44.0 |
| | | 3.1 | 0.34 | 0.79 | 0.0 |
| **Oxide reference** | α-Fe (measured) | 100 | 0.00 | 0.00 | 33.0 |
| | $Fe_3O_4$ (measured) | A: 40 | 0.26 | -0.02 | 48.9 |
| | | B: 60 | 0.67 | 0.00 | 45.9 |
| | $Fe_3O_4$ | A: 30 | 0.27 | 0.00 | 48.8 |
| | | B: 70 | 0.66 | 0.00 | 45.8 |
| | nonstoichiometric-$Fe_3O_4$ (ref 4) | A: 50 | 0.28 | 0.00 | 49.1 |
| | | B: 50 | 0.65 | 0.00 | 45.8 |
| | β-$Fe_2O_3$ | 100 | 0.37 | 0.75 | 0.0 |
| | FeO (measured) | A: 0 | 0.00 | 0.00 | 0.0 |
| | | B: 60 | 1.05 | 0.13 | 0.0 |
| | | C: 40 | 0.91 | 0.63 | 0.0 |
| | nonstoichiometric-FeO (ref 5) | A: 21 | 1.07 | 0.00 | 0.0 |
| | | B: 62 | 1.03 | 0.18 | 0.0 |
| | | C: 17 | 0.94 | 0.44 | 0.0 |

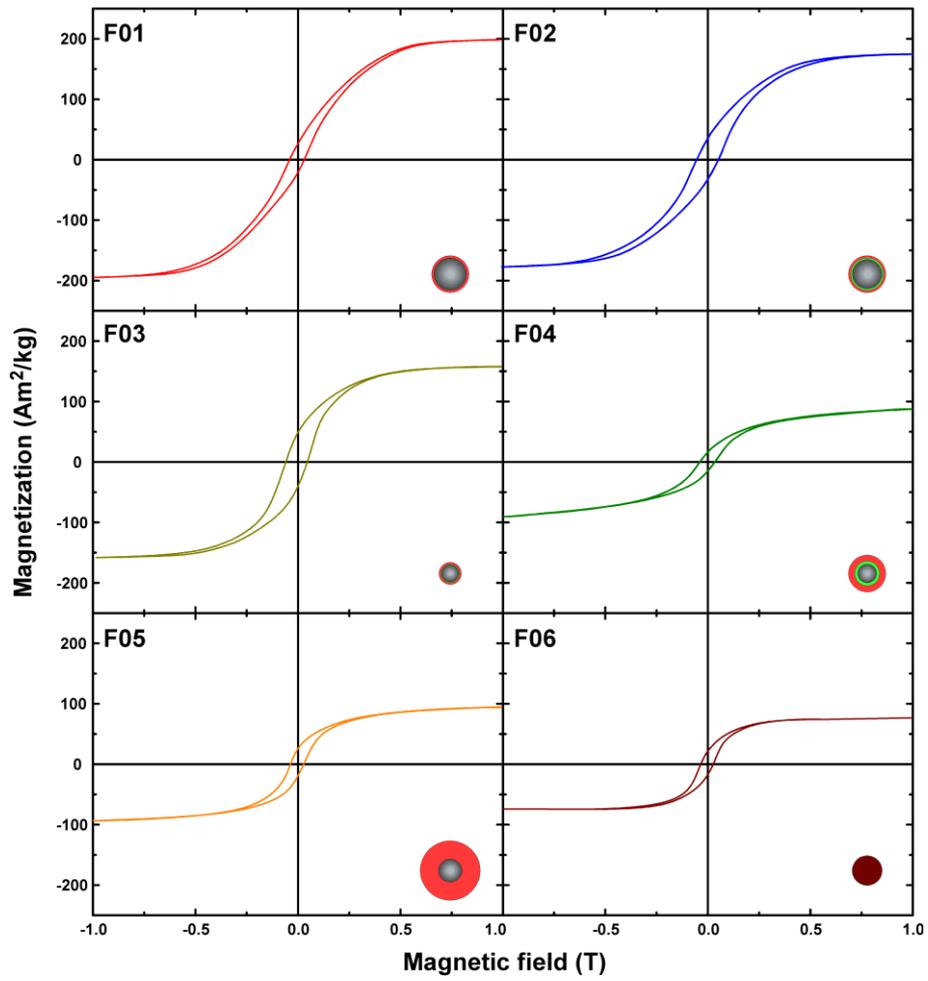

**Figure S5.** Room-temperature hysteresis loops recorded at maximum field of 1 T for samples under study, with decreasing Fe content (F01-F06).

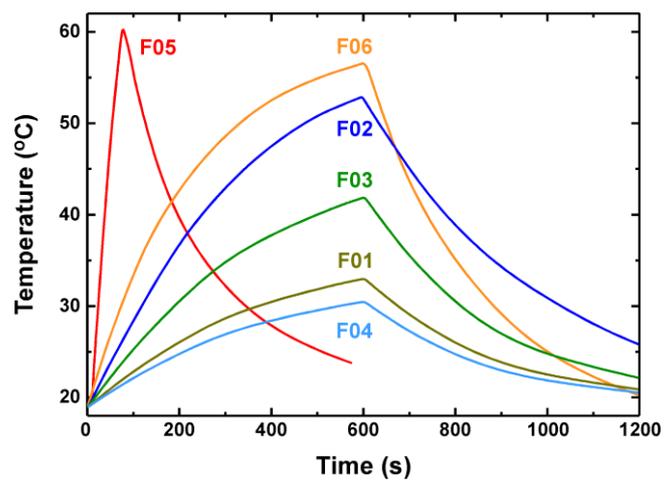

**Figure S6.** Representative hyperthermia curves for dispersions (concentration 2 mg/mL) of samples under AC field conditions of 765 kHz and 30 mT.

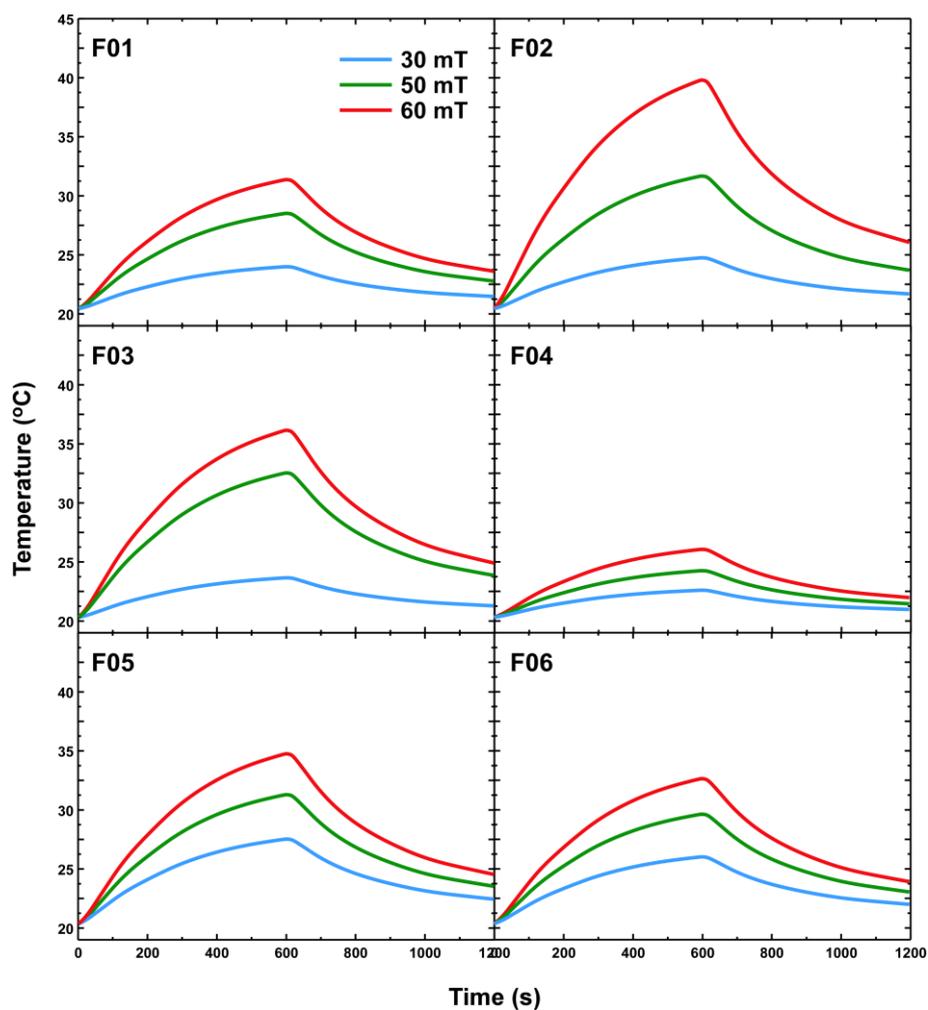

**Figure S7.** Representative hyperthermia curves for all samples (concentration 4 mg/mL) at increasing AC field amplitude (30, 50, 60 mT) for a fixed frequency of 210 kHz.

**MICROMAGNETIC SIMULATIONS**

From the results shown in main text Figure 4 it does not seem to exist any direct correlation between the characteristics of a given core/shell particle (summarized in Table 1) and the released heat. Such apparent heterogeneity in the behaviour includes also the response to different field conditions: note that depending on the field amplitude and frequency, different samples stand for the largest heating.

Seeking to shed light on this, we describe below a technique for computing the dynamic heating capabilities corresponding to the hysteresis losses (i.e. quasi-static hysteresis loops areas times the frequency) of the samples described above. We show that the diagnostics of the particles structural parameters used in the model highly affect switching modes in the magnetic domain. Unless otherwise stated, we neglect dipole-dipole interactions and take into account only exchange, Zeeman and anisotropy contributions.

**S.1 Core/shell parameter dependences**

First, let's turn our attention to the magnetization state of an *isolated only core* particle for a moment. This is traditionally determined by considering the balance between its magnetostatic and exchange energy. The evolution with its size (exploratory factor 1) is usually thought to be from superparamagnetism to blocked single domain to multidomain structure. But Néel pointed out that with increasing sample size, inhomogeneous states become energetically favourable: this is the curling or vortex state. Coey demonstrated such a curling is favoured in materials with high magnetization, details depending on the strength and character of its anisotropy (e.g. they are expected to form in Fe particles with sizes above 22 nm).[6] The first direct observation of such a topologically non-trivial spin texture was made in 1980 using transmission electron holography and electron diffraction.[7] However, these predictions neglect key factors that influence vortex formation, including interactions between neighbouring particles.[8,9] Furthermore, when two magnetic materials are in direct contact, the influence of the proximity effects (factor 2) can for example give rise to mutual imprinting of domain structures, even in the case of materials with different directions of anisotropy. For instance, it was previously shown that the curvature of the underlying surface leads to a coupling between the localized out-of-surface constituent with its delocalized in-surface structure.[10] This stems from the fact that the geometry of a spherical shell prohibits an existence of spatially homogeneous magnetization.[11] Therefore we expect that as soon as the magnetic core is wrapped by another shell material, the vortex stability will be affected.[12] For example, it has been previously shown that for small core-to-shell proportions the single-domain limit comes very close to the size at which the vortex collapses. This collapse represents a second order phase transition.[13] This is accompanied by a maximum remanence that can have a significant influence in the hysteresis losses.

The second aspect to consider is the relatively small heating power in relation to the particle sizes, which might be indicative of non-coherent magnetisation reversal mechanisms; otherwise the heat release would be much higher[14]. Such characteristic strongly reinforces the idea that the behaviour of the experimental samples cannot be interpreted in terms of single-domain particles with coherent rotation of the magnetic moments, as often assumed in theoretical models dealing with hyperthermia properties[15–17]. In addition, the large sizes of the samples point out to a micromagnetic technique as the only feasible approach[18,19] (atomistic procedures would be more accurate but today's computers are unable to simulate the experimental scenario).

So, in our case we illustrate the different roles of the particle size (factor 1) and the intraparticle exchange coupling (factor 2), controlled by the core/shell ratio, which implicitly allows for angular and magnitude dispersion of the anisotropy axes of the shell, and thickness of a paramagnetic interlayer spacer (factor 3). All three characteristics of our approach imply that we can produce high detail magnetization representations using just a few experimental facts and a fraction of the time that current techniques now require.

*Total particle size*

When studying the role of the total particle size of a core/shell particle, the first questions that come to one's mind are, i) how to set up the easy axes of both phases, and ii) which $t_S/r_C$ value(s) to choose. For the sake of simplicity we have first chosen the easy axes of both phases pointing along the same direction (as indicated in main text Table II). Figure S8 compares hysteresis loops of samples with increasing size for the same ratio of $t_S/r_C$ = 1.6 that resulted experimentally in maximum SAR for the field amplitude of 30 mT.

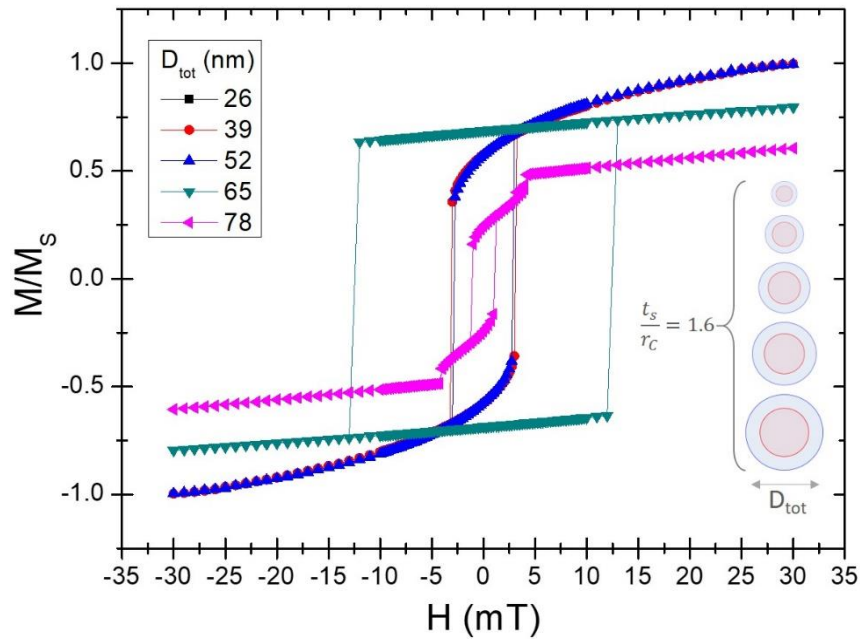

**Figure S8.** Some examples of M(H) curves for a core/shell ratio of $t_S/r_C$=1.6, and different total size, $D_{tot}$. The sketches illustrate the examples in main text Figure 6A.

Obviously, although the simulated magnetization reversals correspond to an idealized particle, valuable information can still be obtained: it seems there is a threshold of the particle size (beyond 52 nm) at which the hysteresis area increases (here shown for the case of 65 nm) until further size enlargement (here shown for 78 nm) is accompanied by a change in the shape and narroweness of the M(H) loops. Seeking a more detailed view on this phenomenon, Figure S9 exemplifies contributions for the core and shell phases separately. Please also note the FM-like coupled behaviour of core and shell.

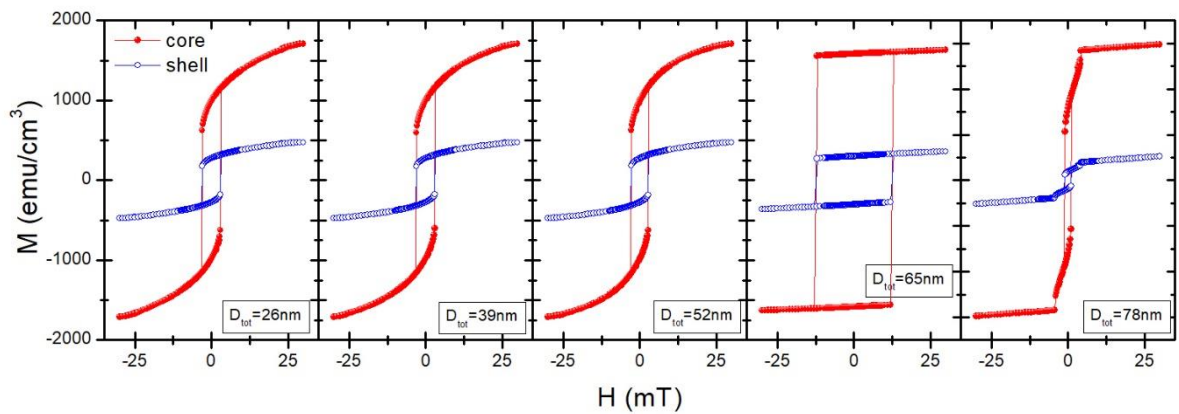

**Figure S9.** Separate contributions of the core and shell magnetization loops upon the increase of particles size for a core-to-shell thickness of 1.6.

Figure S10 analyses the magnitude of the energies involved in the Fe/Fe$_3$O$_4$ core/shell system when an external magnetic field is applied. The results indicate that the smaller-size cases (at least up to 52 nm) undergo a coherent-like reversal (hence the continuous variation of all the three magnetisation projections), with a very unfavourable arrangement for magnetostatic energy. Then, increasing the size favours the appearance of closed flux states that can be easily tailored by varying the particle diameter: magnetization reversal mechanism proceeds *via* domain-wall formation and propagation (irreversible) for the 65 nm case, but continuous (reversible) variation for the 78 nm particle.

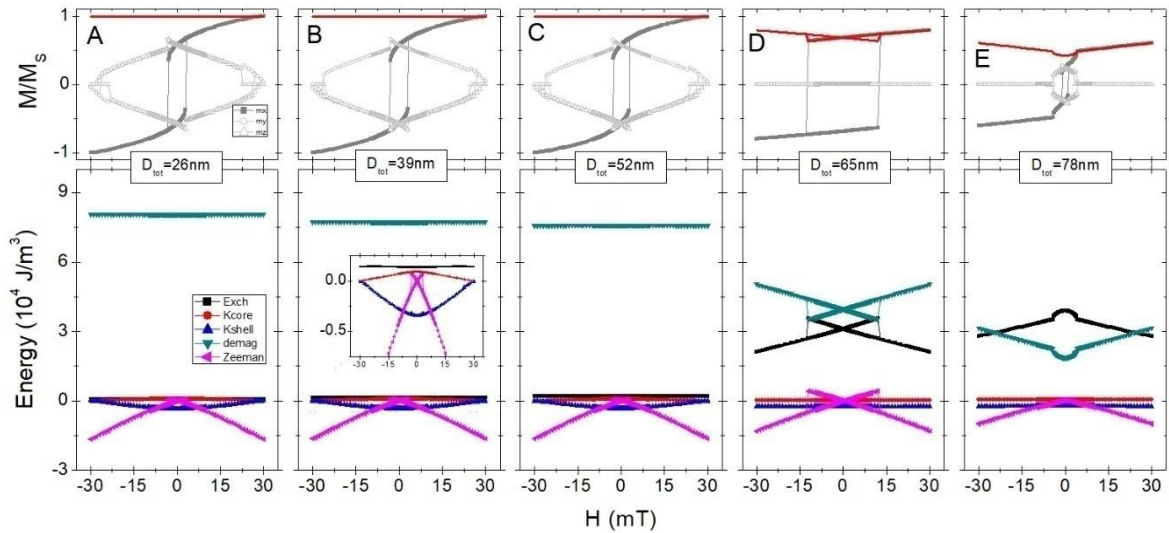

**Figure S10.** Top panels show the (x,y,z) magnetisation components of the same samples as in Figure S9; the red curve stands for the magnetisation module that accounts for irreversibility. The bottom panels show the corresponding energy contributions: exchange, E$_{exch}$; anisotropy of the core (K$_{core}$) and shell (K$_{shell}$); demagnetisation; and Zeeman.

*Core/shell ratio*

We have previously seen that for a given core/shell ratio, the magnetic and hyperthermia properties vary significantly depending on the absolute particle size. In the following we will focus on the issue of varying the core/shell ratio while keeping constant the particle size, as exemplified by the case of a 40 nm particle in Figure S11.

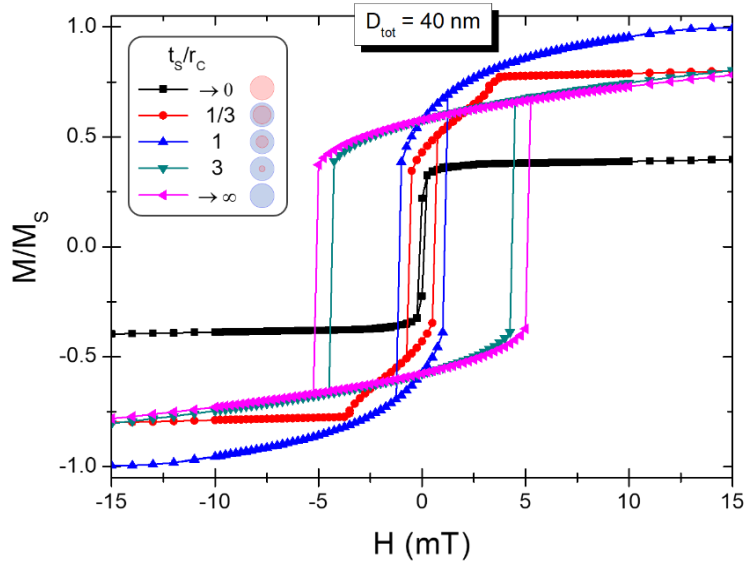

**Figure S11.** M(H) curves for different core/shell particles of same total diameter 40 nm.

The figure summarizes different behaviours that are intermediate between those two extremes of negligible shell (i.e. limit $t_S/r_C \to 0$, only Fe nucleus) and only shell ($t_S/r_C \to \infty$ case, i.e. entire $Fe_3O_4$ particle).

Analogously to what was done in Figure S. 10 we perform next the analysis of the governing energies. The results are shown in Figure S12, together with the magnetisation module and (x,y,z) components: top panels suggest complex behaviours, ranging from non-coherent magnetisation reversal at low $t_S/r_C$ values to coherence developing at values above $t_S/r_C = 1$, which clearly correlates with those seemingly disparate energies in bottom panels.

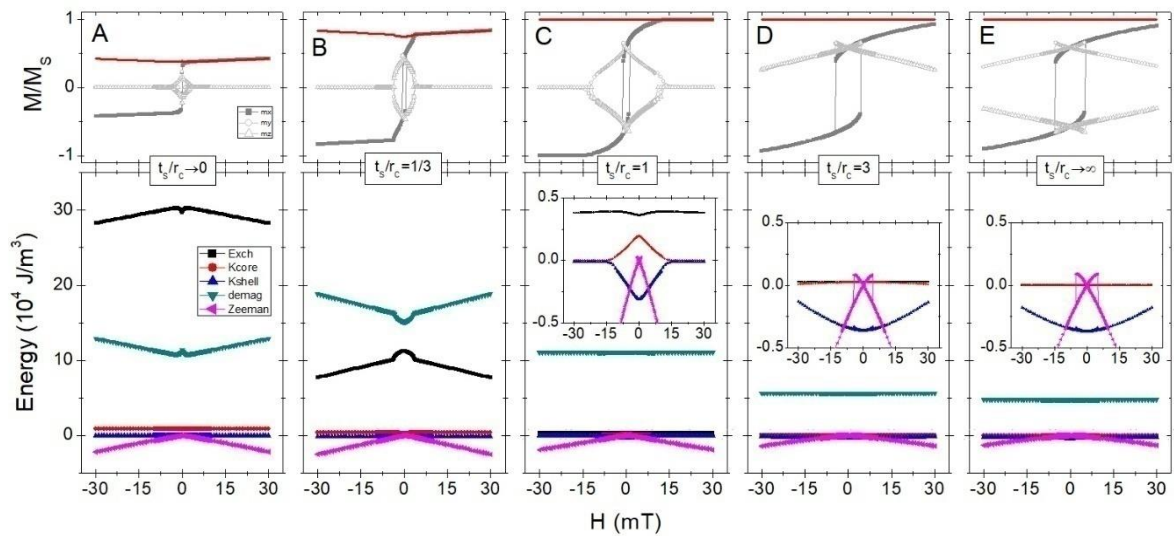

**Figure S12.** Same as in Figure S10, but now for the different core/shell-ratio cases of main text Figure 6B.

Complementarily, Figure S13 shows the contributions from core and shell magnetization separately.

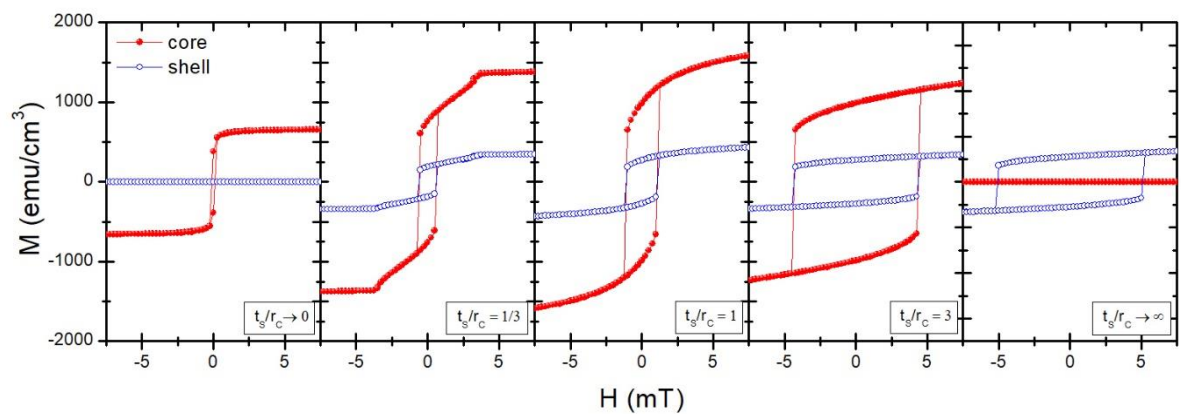

**Figure S13.** Loops depicting the core and shell contributions to 40 nm particle's magnetization upon the increase of the shell thickness.

*Thickness of interlayer spacer*

So far, we have observed a strong influence of both the total particle size and core/shell ratio. Nonetheless, a close examination indicates that even considering similar values of both parameters, the SAR values may differ; note for example that despite samples F01 and F02 do have very similar sizes about 50 nm and close $t_S/t_C$ ratios their heating performances are different, see Figures S6 and S7. We recall that the only difference that can at all be detected between these two samples is in the shell composition. Thus we generalize the analysis to include a separating FeO spacer. Under laboratory conditions, the FeO layer is expected to behave as a paramagnet, consequently we have treated such distinct behaviour by assigning a very low $M_S$ (see main text Table 2). Some representative M(H) hysteresis loops are shown in Figure S14.

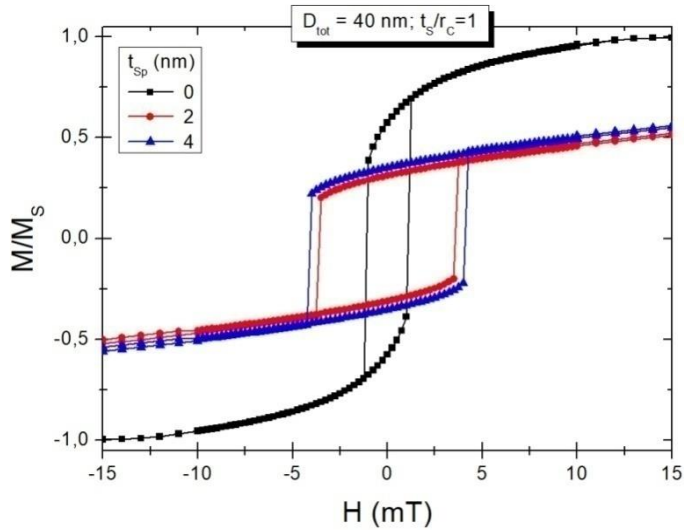

**Figure S14.** M(H) curves for a fixed absolute particle size of 40 nm, and roughly the same core-to-shell aspect ratio (about 1), for different thicknesses of the separating FeO spacer.

Noteworthy, the results plotted in Figure S14 indicate a radical change of the particle behaviour when a nonmagnetic interlayer is present, with a hardening and significant lowering of the total magnetization that strongly points to a non-coherent magnetisation arrangement. In order to check this, we analyse the magnetization components and energies, summarized in Figure S15: on the one hand, the hysteresis loops are non-exactly symmetrical; on the other hand, while some of the energy features resemble those observed in Figure S12, the fact is that the magnetisation evolution is completely different. Note, for example, that while the case of Figure S12C is represented again in Figure S15A, the features in the energy panels S15B-C correspond to completely different non-coherent magnetisation switching..

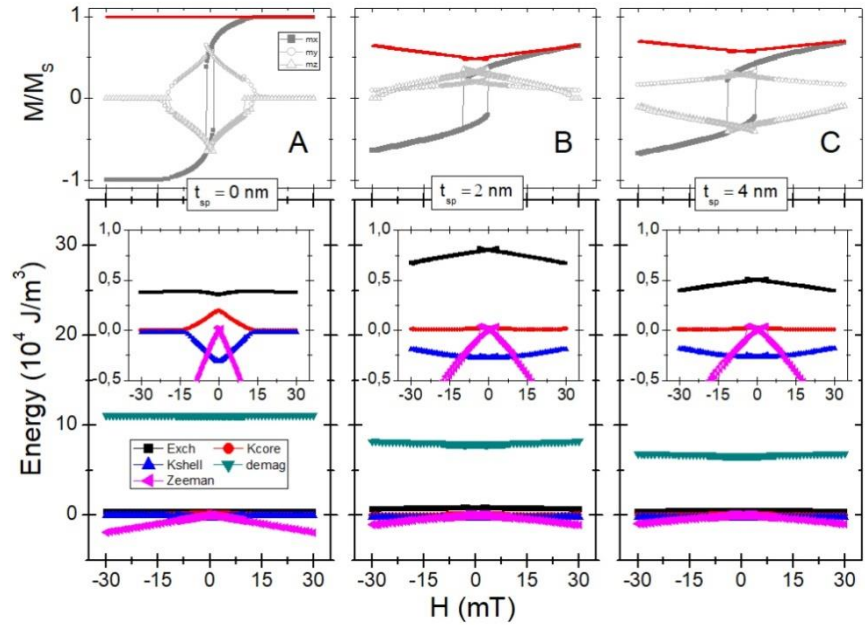

**Figure S15.** Top panels show the (x,y,z) magnetisation components for the different core/shell-ratio cases of Figure S14, together with the magnetisation module (red curve). The bottom panels show the corresponding energy contributions. The labels are the same as above.

In order to gain further insight into the observed behaviour and in analogy to what has been done in the previous sections, Figure S16 shows the separate contributions of core and shell. Now, when a nonmagnetic interlayer is included, a completely different scenario emerges since the core and shell phases couple AFM-like. Remarkably, this antiparallel coupling leads to broader hysteresis losses, thus offering an additional way to tune the particles heating response (see Figure S7).

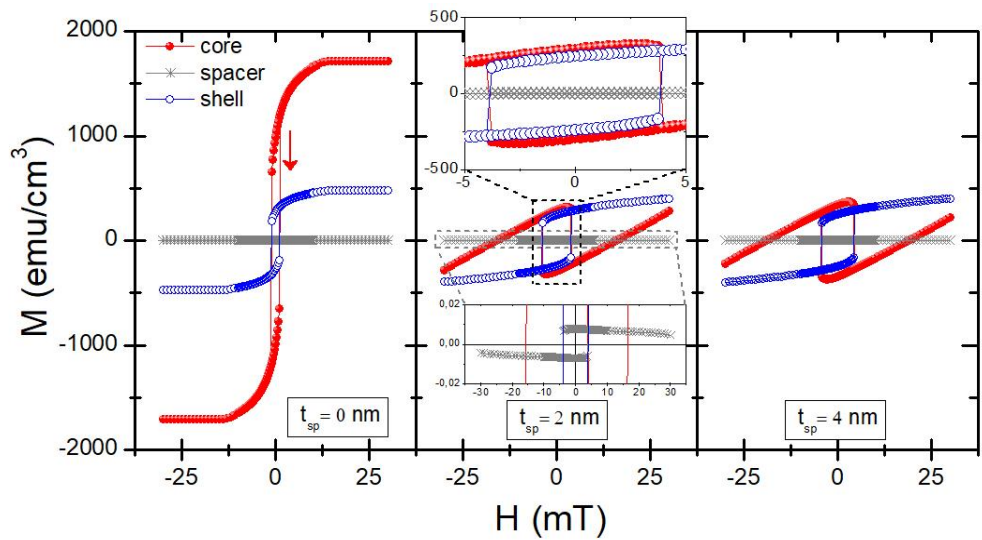

**Figure S16.** M(H) hysteresis loops separating the core and shell contributions to 40 nm particle's magnetization upon the inclusion of an FeO nonmagnetic interlayer. The inset amplifies the region of the switching field.

*Shell crystallites' sizes*

Although we have assumed up to now that both core and shells were single-crystal phases, it is well-known that native oxide layers are polycrystalline and that such nanostructure may significantly modify the magnetic response of the particles[20]. Accordingly, this was solved using a mesh with different domain-size. The results are shown in Figure S17.

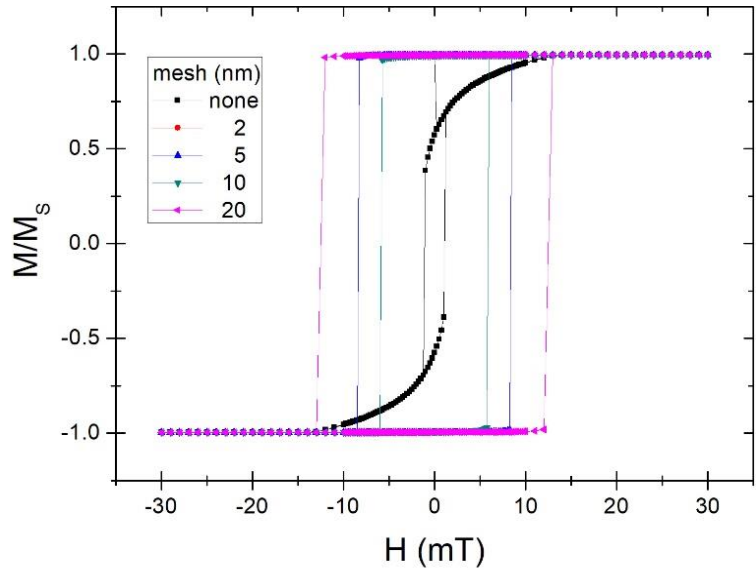

**Figure S17.** Mesh-size dependent effect of shell's grain size on magnetic properties M(H) for a crystalline/polycrystalline Fe/Fe$_3$O$_4$ particle of 40 nm diameter and $t_S/r_C$ = 1.

Dependent on the refinement of model mesh, a radical change of the hysteresis loops shape -from a smooth magnetisation change to an abrupt square one- accompanied by a large increase of both coercivity and remanence, is observed. In this particular case it would lead to a huge increment of the dissipated energy *via* Néel reversal. Further insight into the process is thus provided by Figure S18, from which it follows that:

- The reversal process is always coherent.
- In comparison with the homogeneous-shell case, in which core and shell anisotropy energies have opposite signs at low fields, the presence of a polycrystalline shell sets the anisotropy of the core at basically zero, while that of the shell falls to negative values.
- Also in contrast to the single-crystal shell case, for which the exchange energy was always the dominating (positive) one, in the polycrystalline shell particle the Zeeman energy overcomes the exchange factor at some field value, when the sudden magnetization reversal occurs.

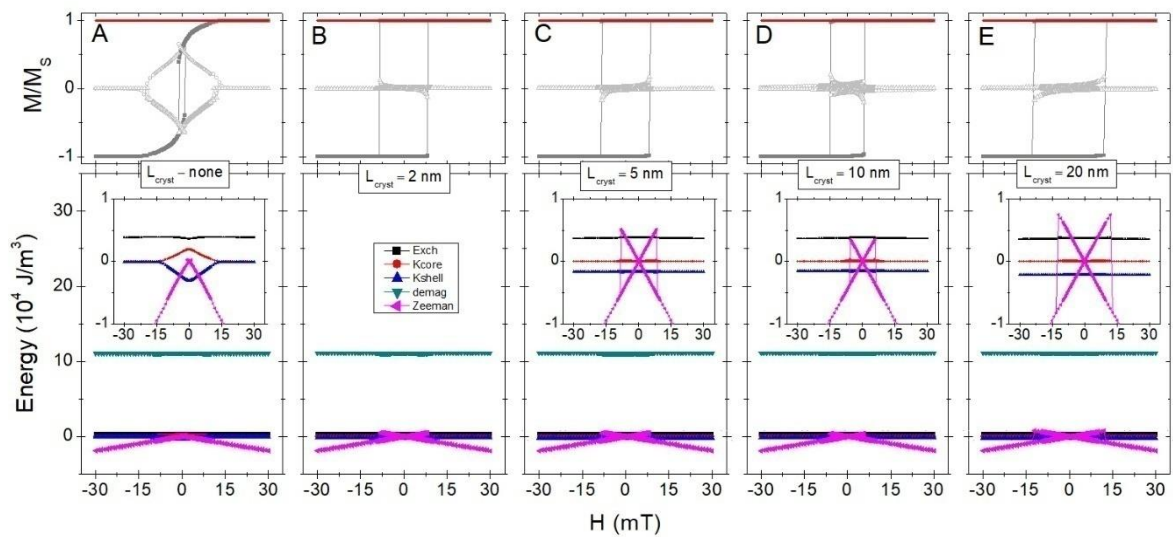

**Figure S18.** The labels are the same as in Figure S15, for a particle discussed in Figure S17.

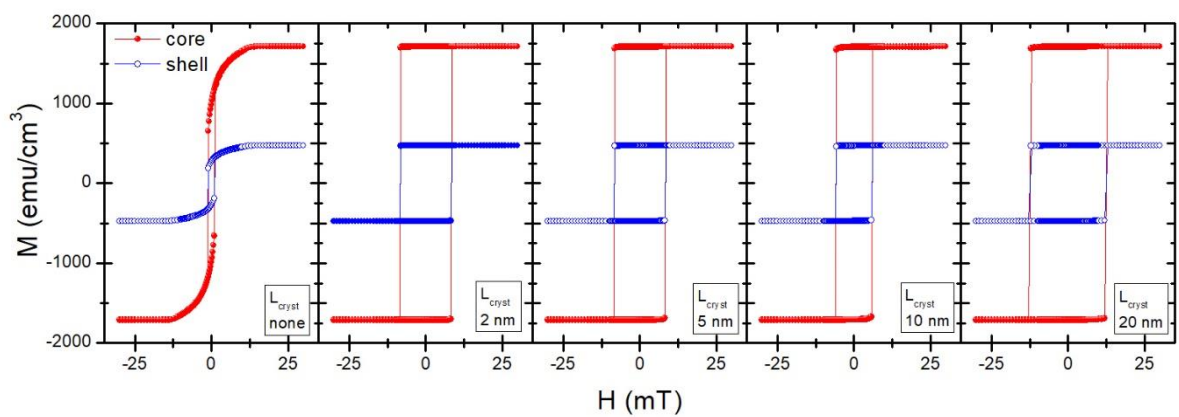

**Figure S19.** Core and shell contributions to magnetization of a 40 nm particle composed of a crystalline 20 nm Fe core and a polycrystalline 20 nm shell made of magnetite, for various average crystallite sizes.

The results shown in Figure S19, with the separate magnetic evolution of the core and shell phases, corroborates the previous observations on the coherent nature of the magnetisation reversal, and the dramatic change of the hysteresis loop shape to a fully squared one, i.e. equivalent to optimum heating process.

## S.2 Angular-dependent properties and polycrystalline shell

An important characteristic to consider when simulating magnetic materials is the distribution of the magnetocrystalline easy axes. A numerical error is unavoidably introduced by the effect of surface roughness arising from the discrete nature of the cubic lattice cells, as discussed in Refs.21. Consequently, in order to avoid possible nonphysical effects the simulated curvilinear surface must be smooth enough, which requires a rather detailed mesh; hence, the modelling becomes very expensive from a computational point of view.[22] Thus, we introduce a second mesh that accounts for randomly distributed easy axes of the shell, as sketched in Figure S20. Note that different lattice cell sizes have been used in the calculation depending on the size of shells.

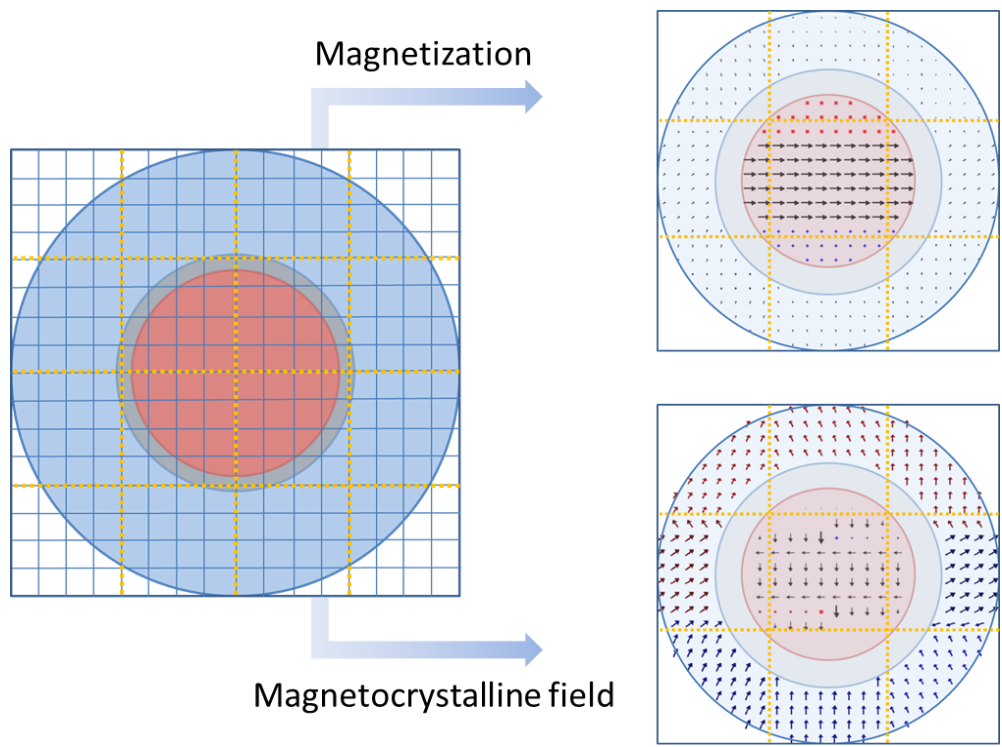

**Figure S20.** Left: Sketch illustrating how a second mesh (dotted yellow parallel lines) is superimposed over the magnetic one (thin blue parallel lines) and used to build shell crystallites with different directions of the magnetocrystalline anisotropy. Right: top figure shows the magnetic configuration at remanence, associated to the bottom nanoparticle with anisotropy axes distributed in many directions within the plane of the shell's surface.

The effect of the polycrystalline shell regarding the magnetic response is summarized in Figure S21, which compares a $Fe/FeO/Fe_3O_4$ core/spacer/shell particle (F01) with a homogeneous "only-shell" $Fe_3O_4$ particle (F06).

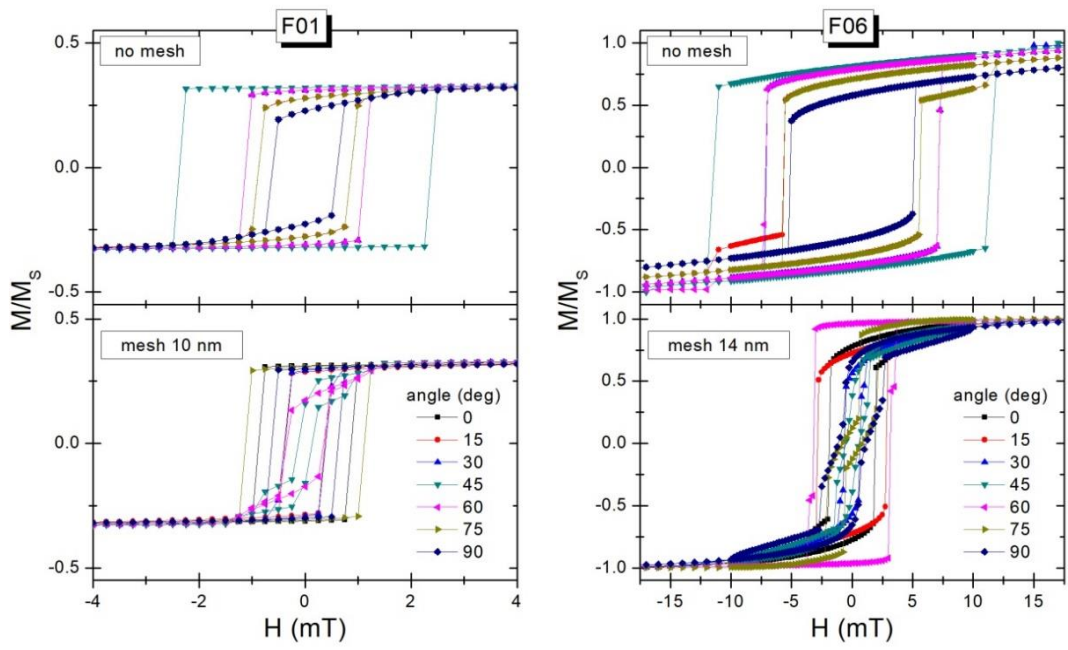

**Figure S21.** Effect of grain size distribution on magnetic properties. Left: Top panel shows the angular-dependent M(H) loops corresponding to sample F01 with a single-crystal shell, whereas the bottom panel corresponds to the same sample with a shell anisotropy mesh of 10 nm. $\theta = 0$ accounts for the field along the [100] easy axis of magnetization of iron. Right: idem, but for sample F06 with a mesh of 14 nm.

It is observed that the presence of randomly distributed crystallites in the shell reduces significantly the angular-dependent properties. This feature provides confidence on the rationality of our simulations despite using only one of the available field directions. In addition, Figure S22 confirms that the presence of many shell crystallites decreases the hysteresis losses and the rotational symmetry around at 45 degrees.

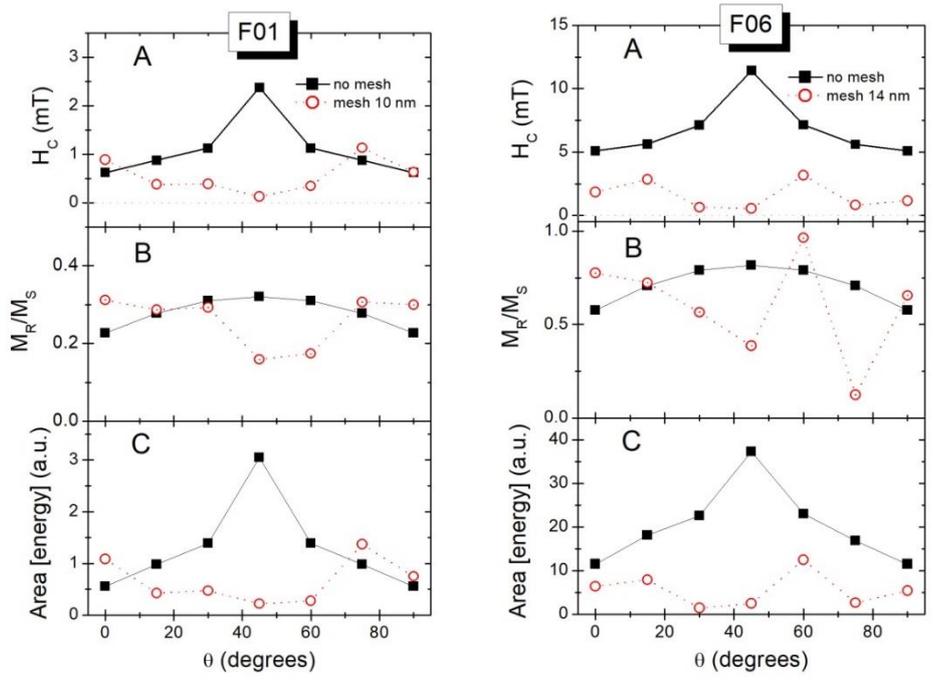

**Figure S22.** Room-temperature dependence of coercive field (A), remanence ratio (B) and hysteresis loop area (C) on the applied field direction for samples F01 (left panels) and F06 (right panels).

## S.3 Simulation of the experimental samples

Building on the above framework, next we benchmark our experimental results against modellizations. The magnetisation results are shown in Figure S23.

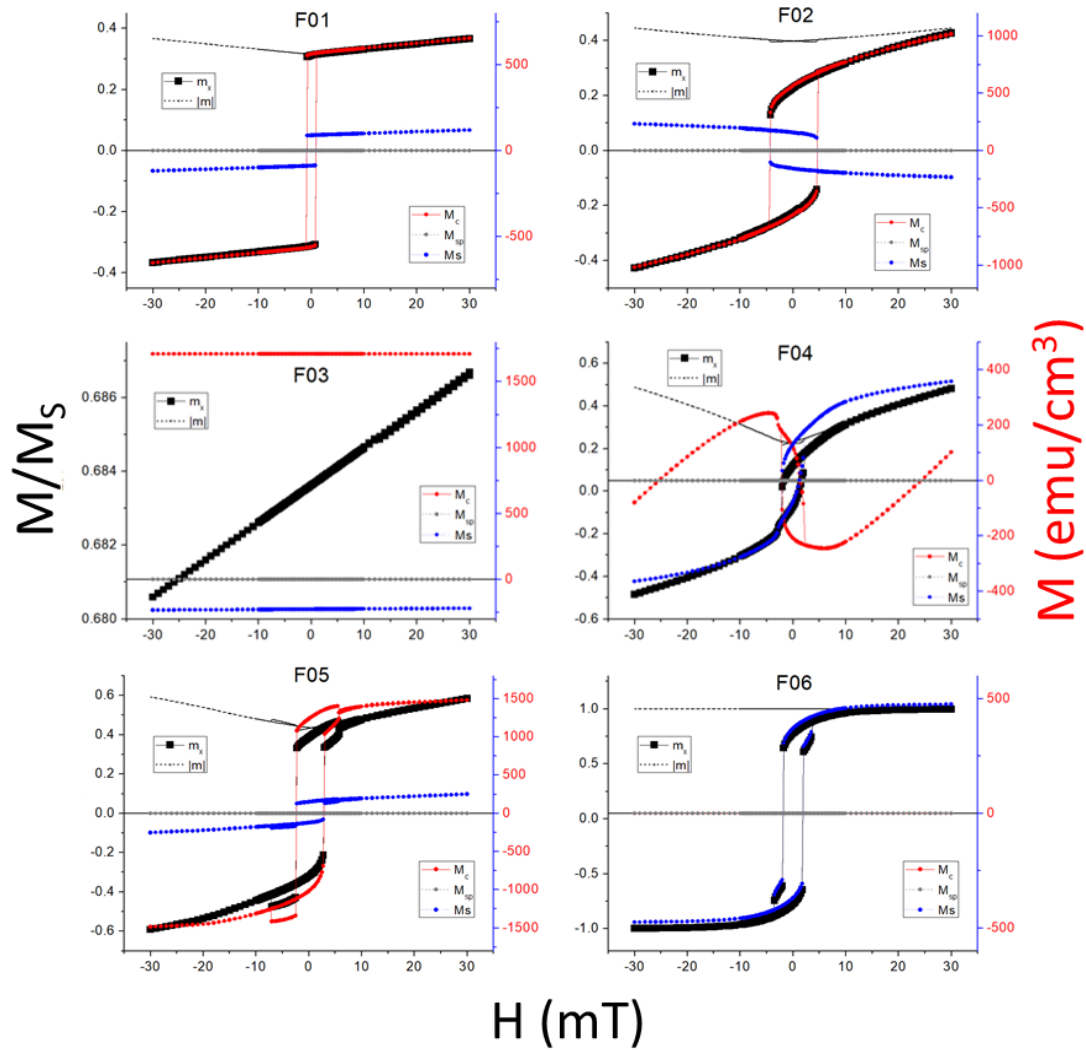

**Figure S23.** Micromagnetic simulation of the magnetisation M(H) loops of the six experimental samples summarized in main text Table 1. in quasi-static field conditions up to 30 mT. In each case, two different type of curves are shown: magnetisation of the entire particle along the field direction ($m_x=M_x/M_S$) and module of the magnetisation ($|m|=|M|/M_S$), in normalised units (left vertical axis); and core ($M_c$), shell ($M_s$) and paramagnetic interlayer ($M_{sp}$) contributions to magnetization along the field direction, in real units (right vertical axis).

It is noted that:

- all samples the system exhibits a hysteretic loop except for sample F03, which seems to be under minor-loop conditions.
- samples F01, F02, F04 and F05 undergo a coherent-like reversal pointed out by the |m| variation, whereas it remains essentially unchanged for samples F03 and F06.
- samples F01 and F05 exhibit FM-like coupling between core and shell, whereas F02, F03 and F04 exhibit AFM-like arrangement (note that sample is a single-phase).
- while for samples F01, F02 and F06 the magnetisation follows a symmetric path, in samples F04 and F05 the positive-to-negative and negative-to-positive field branches are clearly asymmetric (note that nothing can be said about sample F03 ).

These main aspects could certainly be related to key magnetic-hyperthermia aspects (for example, having FM-like coupling would be in principle more desirable because it facilitates bigger loop areas), but the fact is that there is only one sample (F03) showing a clearly divergence between the theoretical predictions and the measured values. A detailed analysis of this sample shows that the second-order angular-dependence assumption does not work for this particular case because parallel-orientation is markedly different (see Figure S24), despite the fact that in general the particle undergo major-loop conditions. Remarkably, the squareness of the loops strongly promises quite a large heating capability.

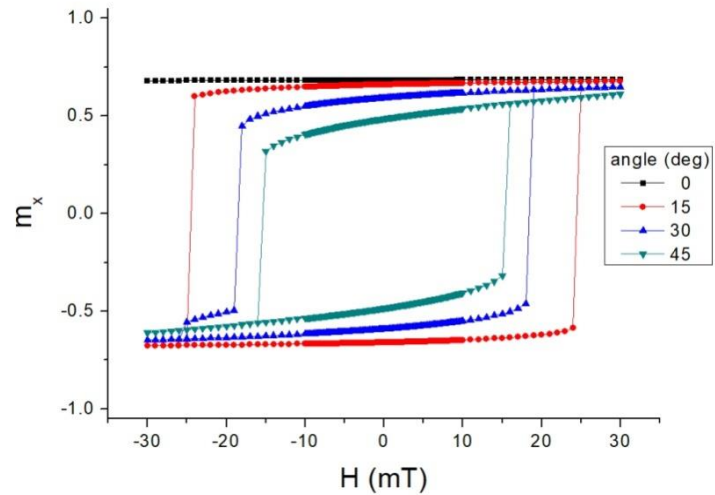

**Figure S24.** Some angular-dependent hysteresis M(H) loops corresponding to sample F03.

Thus, we further analyzed the magnetic response of sample F03 under higher field amplitudes. The results are shown in Figure S25. Noteworthy, the typical Stoner-Wohlfarth behaviour is recovered, with the particularity of being the cycle the resultant of the coherent response of two strongly-coupled lattices in which the core is leading (in contrast to sample F04, for example, where the shell dominates the magnetic switching). In conclusion, by increasing the field amplitude this particular sample F03 may exceed other samples, in qualitatively agreement with the experimental results depicted in main text Figure 4.

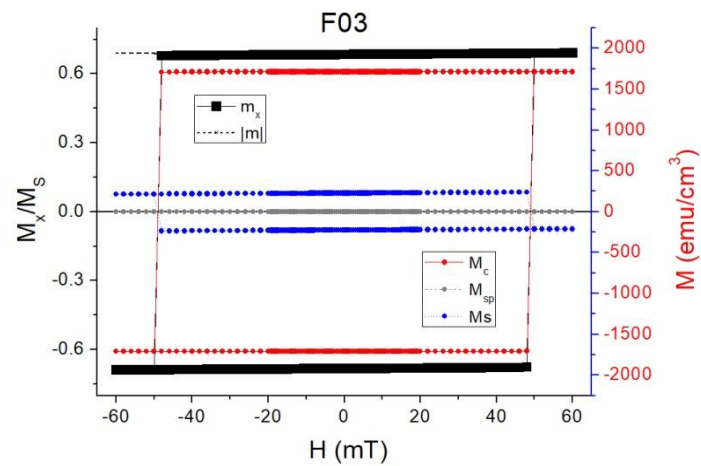

**Figure S25.** Magnetization versus field loops for sample F03, same as in Figure S23 but now increasing the applied field up to 60 mT.

Quantitative comparison of the heating efficiencies is made possible after considering a linear response of the hysteresis loops with the applied frequency. This is shown in Figure S26 that reproduces main text Figure 7 except for sample F03. Yet, the simulation clearly overestimates the experimental measurements in this sample.

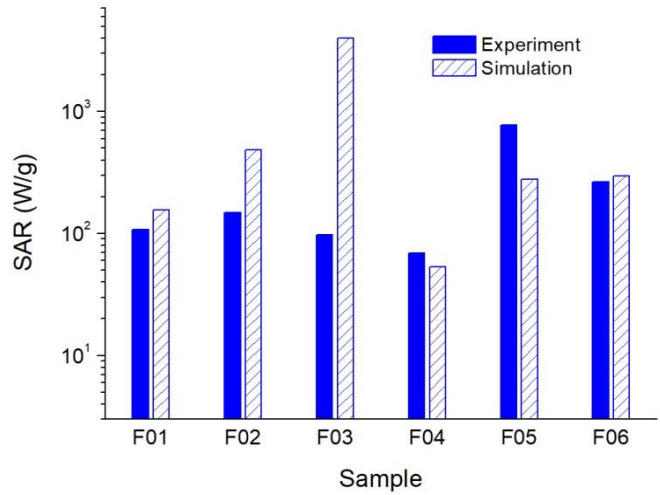

**Figure S26.** Comparison between experimental SAR values (filled blue bars) and theoretical predictions obtained based on non-interacting conditions and an alternating field of 765 kHz and 30 mT. Note the logarithmic scale. The results are the same as those plotted in main text Figure 7 except for sample F03, for which the SAR was estimated from the weighted M(H) average of a randomly distributed population of angles between the field and easy axis.

**S.4 Trial model modifications**

To check whether some key ingredient might be missing in the assumed model, some alternative possibilities were tested, mainly related to including surface anisotropies (at the outer layer of the shell, and/or at the interfaces), changing the character of the separating interlayer (to nonmagnetic), or including dynamical effects (results not shown). The results are summarized in the following figures.

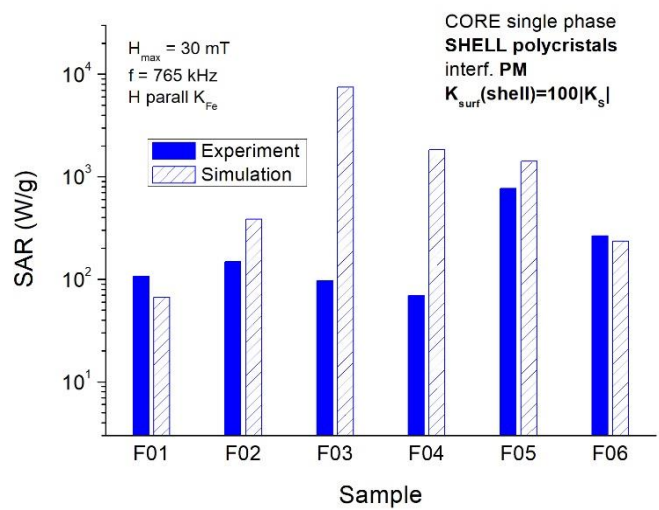

**Figure S27.** Same results as shown in Figure S26, but considering a perpendicular surface anisotropy on the outer layer of the shell, of magnitude 100 times that of the shell.

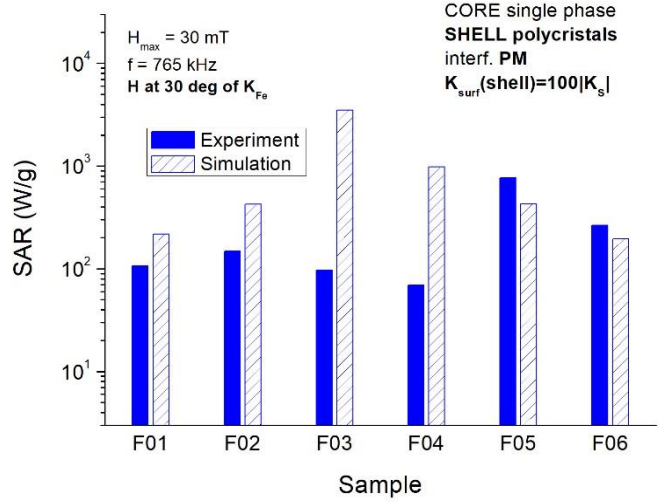

**Figure S28.** Same as previous Figure S27, just the field applied at 30 degrees from the Fe easy axis.

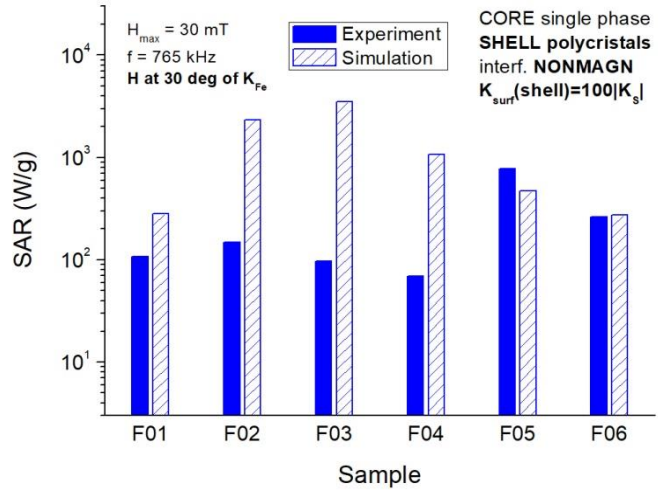

**Figure S29.** Same as previous Figure S28, just treating the interlayer as a nonmagnetic spacer.

None of the above offers better results. Including other modifications as e.g. different perpendicular surface anisotropies either at the outer surface of the shell, and/or at the interfaces, always resulted in a worst comparison to the experiment. The same occurs in general with other modifications as e.g. change in the shell mesh defining the average shell crystallites, or including a mesh at the core (for possible crystals at the core too). Finally, it is also worth to mention that while a complete dynamical simulation is computationally unfeasible, and beyond of the scope of this paper, preliminary results (not shown) indicated that by no means either thermal fluctuations (at room temperature) or dynamical effects would lower the SAR estimation from sample F03.

**S.5 Role of particle-particle dipolar coupling on the heating performance of sample F03**

In any magnetic material, there are preferential directions along which magnetic moments find it easier to align; these directions are known as the easy axes. But more complicate issues arise in the presence of dipolar interactions and assemblies,[23] existing a coupling between the dipolar interaction

-23-

field and the alignment of the nanocrystal easy axes.[24] We refer the reader to our former papers [25,26] for an insight into the effects of nanoparticle aggregation on the heating properties.

Therefore, we have tentatively explored the influence of interparticle dipolar interactions on the heating performance of sample F03 in a last effort to meaningfully explain the experimental data. In Figure S30 we consider the effect of a non-switchable particle (i.e. with easy axis closely aligned with the external AC field) that projects a fixed bias field onto another close by. Obviously, the strength of such bias field is proportional to the interparticle distance and their relative orientation.

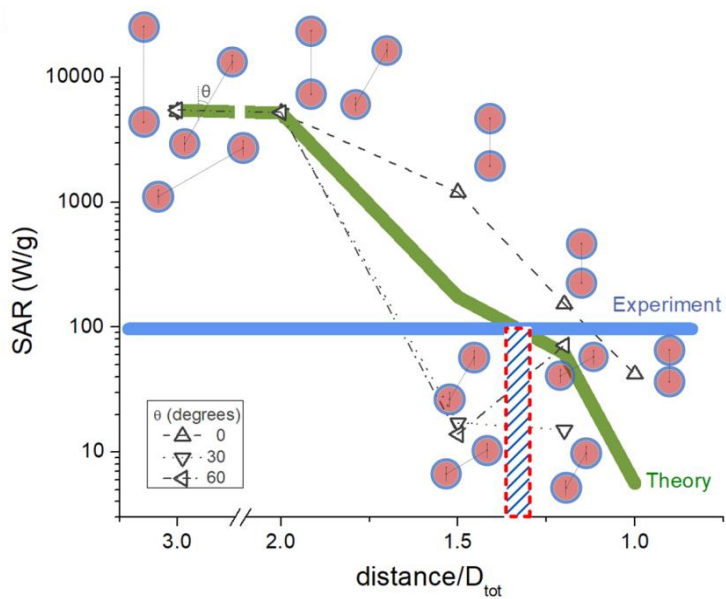

**Figure S30.** SAR values of a F03 particle under the effect of another similar non-switching particle that acts upon creating a local bias field, of different strength depending on the interparticle distances (x-axis, in units of the total diameter of a F03 particle) and arrangements: parallel-aligned particles (0 degrees), at 30 or 60 degrees, as illustrated by the sketches. The green line stands for the weighted average; horizontal blue thick line indicates the experimentally measured SAR, and the dashed red square reproduces that in main text Figure 4.

The results indicate a decrease of the heating performance with increasing interparticle coupling, thus supporting our interpretation. The effect is enhanced for interparticle distances smaller than twice the particle diameter, and which closely matches the experimental data in main text Figure 4 measured by calorimetry. This level of agreement between measured dynamic SAR and estimated quasi-static hysteresis losses times the frequency is remarkable considering the small number of coefficients that need to be determined and that these were independent experiments.